\documentclass[pra,twocolumn,showpacs,aps,amssymb,nofootinbib]{revtex4}
\usepackage{amssymb}

\usepackage{graphicx}
\usepackage{amsmath}
\usepackage{times}
\begin{document}
\title{Quantum Kinetic Theory of BEC Lattice Gas:Boltzmann Equations from 2PI-CTP Effective Action}
\author{$^{1,2}$Ana Maria Rey, $^{1}$B. L. Hu,$^3$, Esteban Calzetta and  $^{2}$Charles W. Clark}
\date{\today}
\affiliation{$^{1}$ Department of Physics, University of Maryland,
College Park, MD 20742}
 \affiliation{$^{2}$National Institute of Standards and Technology, Gaithersburg, MD
20899, USA.}
 \affiliation{$^{3}$ Departamento de Fisica, Facultad de Ciencias Exactas y
Naturales, Universidad de Buenos Aires- Ciudad Universitaria, 1428
Buenos Aires, Argentina}

\begin{abstract}
We continue our earlier work  [Ana Maria Rey, B. L. Hu, Esteban
Calzetta, Albert Roura and  Charles W. Clark,  Phys. Rev. A 69,
033610 (2004)] on the nonequilibrium dynamics of a Bose Einstein
condensate (BEC) selectively loaded into every third site of a
one-dimensional optical lattice. From the two-particle irreducible
(2PI) closed-time-path (CTP) effective action for the Bose-
Hubbard Hamiltonian, we show how to obtain the Kadanoff-Baym
equations of quantum kinetic theory.  Using the quasiparticle
approximation, we show that the local equilibrium solutions of
these equations reproduce the second- order corrections to the
self-energy originally derived by Beliaev. This work paves the way
for the use of effective action methods in the derivation of
quantum kinetic theory of many atom systems.
\end{abstract}

\maketitle

\section{Introduction}

In many respects, the dynamics of cold atoms in optical lattices
resemble those of electrons in crystals. Cold-atom systems exhibit
many favorable attributes for studying quantum many-body dynamics,
such as the absence of defects in the optical lattice, and the
high degree of experimental control over all relevant parameters
\cite{NISTLattice,Blairla}.  In particular, by varying the depth
of the optical lattice, the superfluid-insulator phase transition
can be induced. For weakly-confining optical lattices, the system
has macroscopic quantum coherence, and interesting matter wave
interference phenomena induced by the periodicity of the lattice
have been demonstrated in experiments
\cite{Greiner2001a,Denschlag2002a,Cataliotti,Morsch2001a}. For
tightly-confining lattices, the matter-wave coherence is lost, and
the system undergoes a transition to the Mott-insulator
phase\cite{Fisher}. This regime has become also experimentally
accessible \cite{Greiner2002a, Orzel2001a, Bloch}. Outside the
weakly interacting regime, standard mean field techniques are
inapplicable to describe the evolution of the system, and
alternative  methods are required.

Motivated by a recent patterned loading experiment
\cite{NISTLattice}, we previously adopted a functional effective
action approach capable  of dealing with non equilibrium
situations that require a treatment beyond mean field theory
(\cite{byHFB}, hereafter I). We applied the CTP functional
formalism \cite{ctp} and the two-particle irreducible (2PI)
effective action \cite{2pi} to the Bose-Hubbard Hamiltonian, and
derived equations of motion. This method enabled us to go beyond
the Hartree-Fock-Bogoliubov (HFB) approximation
\cite{Griffin,Morgan,Bogoliubov} and to incorporate nonlinear and
non-Markovian aspects of quantum dynamics, which underlie
dissipation and fluctuation phenomena.

In its pristine form the 2PI-CTP equations of motion for the mean
field and the two-point correlation function are complicated
nonlocal nonlinear equations, which defy even numerical solutions
for realistic experimental systems with many lattice sites. It is
obvious that to get more physical insight we need ways to simplify
this full theory. In this paper we continue this investigation
with the goal of showing how to formulate a quantum kinetic theory
\cite{Kadanoff,Kane} by way of the 2PI-CTP formalism. For earlier
work addressing this problem in quantum field theory, see
\cite{CH88}-\cite{Berges}.  There exist an extensive literature on
quantum kinetic theory, many addressing BEC dynamics with
condensate-noncondensate interactions \cite{Griffin},
\cite{Millena}-\cite{GarZol}. Those relevant to our present
discussions are \cite{Griffin},\cite{Millena}-\cite{Proukakis}.

Towards this goal, we ask the question when quantum kinetic theory
is a reasonable attainable limit of the more complete theory based
on the 2PI-CTP effective action.  Physically, a kinetic theory
regime exists when the system dynamics has a clear separation of
two time (or length) scales, one pertaining to the macroscopic
scale describing the kinetic motion such as the mean free time and
the other to the microscopic scale such as the duration of
collision event. Alternatively, when perturbations induce
disturbances of wavelength longer than the thermal wavelengths and
frequencies much lower than characteristic excitation frequencies,
standard kinetic theories may give a reasonable description of the
system's dynamics. This is the case for weakly interacting gases
confined by a slowly-varying external potential. For quantum
systems, when the quantum features of the many-body system act
effectively only on the microscopic scale (e.g., when one can use
a quasiparticle type of approximation), quantum kinetic theory can
provide an adequate description. It fails  when such a two-time
separation does not exist, such as in strongly correlated systems
or systems with macroscopic quantum coherence \footnote{We have in
mind systems whose quantum coherence or correlation or
entanglement extends to macroscopic dimensions. Examples are
coherence tunneling phenomena \cite{CalLeg83}, quantum properties
of microelectro-mechanical systems \cite{Schwab,Tian} and of
course, BEC, which certainly has macroscopic quantum coherence.
The impossibility of a two-time separation refers only to the
condensate state alone. The interaction between the condensate and
the non-condensate atoms can under general conditions allow a
two-time separation and a kinetic theory description, as is the
topic of our present discussion and much prior work}.

The organization of this paper with a brief of our findings is as
follows. In Sec.\ref{sec2} we summarize our prior results  for the
HFB and second order equations of motion \cite{byHFB} and express
them with lightened notation in a more compact form. In Sec.
\ref{sec3} we discuss how a quantum kinetic theory can be derived
from a quantum theory of interacting particles. We first discuss
this issue under more general conditions, where a two-time
separation may not exist. A kinetic theory is obtained from the
full hierarchy of correlation functions  by truncation of higher
order correlations and the imposition of causal factorizable
conditions. We use the \textit{n}PI-effective action to illustrate
this conceptual framework. In Sec. \ref{Boltzmann} we  focus on
situations where there is a two-time separation in the system
dynamics. We delineate the physical conditions and show the
procedures in deriving   quantum kinetic equations from the
2PI-CTP equations of motion. Then we introduce further
simplifications and discuss how to derive the familiar Boltzmann
equations. In Sec.\ref{sec5} we study how these kinetic equations
admit, as a special yet important case, equilibrium solutions.  We
show that under the Popov approximation the second order 2PI
equations yield to the same second-order damping rates originally
obtained in Beliaev's pioneered work \cite{Beliaev} but with a
modified effective mass due to the presence of the lattice. In
Sec.\ref{sec6} we conclude with a few general remarks.

\section{2PI-CTP effective action for the Bose-Hubbard Hamiltonian}
\label{sec2}

Here we summarize the structure of this method and collect the
useful equations obtained from our earlier investigation
\cite{byHFB}. We will refer to the numbering of equations therein
with a prefix I.

\subsection{The Bose-Hubbard Hamiltonian}

The one dimensional Bose-Hubbard Hamiltonian is given by

\begin{eqnarray}
\hat{H}&=&-J\sum_{i}(\hat{a}_{i\;}^{\dagger }\hat{a}_{i+1\;}+\hat{%
a}_{i+1\;}^{\dagger }\hat{a}_{i\;})+ \sum_{i} V _{i}\hat{a}_{i}^{\dagger }\hat{a}_{i}\notag\\+&&\frac{1}{2}U\sum_{i}\hat{a%
}_{i\;}^{\dagger }\hat{a}_{i\;}^{\dagger }\hat{a}_{i\;}\hat{%
a}_{i\;} , \label{BHH}
\end{eqnarray}
 where $\hat{a}_{i}$ and
$\hat{a}_{i}^{\dagger }$ (called $\hat
\Phi_i,\hat{\Phi}_{i}^{\dagger }$ in Paper I) \ are the bosonic
operators that annihilate and create an atom on the site $i$.
Here, the parameter $U$ denotes the strength of the on-site
repulsion of two atoms on the site $i$; the parameter $V _{i}$
(called $\epsilon_i$ in Paper I) denotes the energy offset of each
lattice site due to an additional external potential that might be
present (such as a magnetic trap), and $J/\hbar$ denotes the
hopping rate between adjacent sites. Next-to-nearest neighbor
tunneling matrix elements are \ typically two orders of magnitude
smaller than the nearest-neighbor ones, and to a good
approximation they can be neglected. The Bose-Hubbard Hamiltonian
should be an appropriate model\cite{Jaksch} when the loading
process produces atoms in the lowest vibrational state of each
well, with a chemical potential \ smaller than the energy
separation to the first vibrationally excited state.

 The classical action associated with the Bose-Hubbard
Hamiltonian (\ref{BHH}), is given in terms of the complex fields
$a _{i}^{{}}$ and $a _{i}^{\ast }$ by

\begin{eqnarray}
S[a _{i}^{\ast },a _{i}]&=&\int dt\sum_{i}i\hbar a _{i}^{\ast
}\partial _{t}a _{i}+ \int dt\sum_{i}J\left( a _{i}^{\ast }a
_{i+1}+a _{i}a _{i+1}^{\ast }\right)\notag \\&& -\int
dt\sum_{i}V_i a _{i}^{\ast }a _{i}-\int dt\sum_{i}\frac{U}{2}a
_{i}^{\ast }a _{i}^{\ast }a _{i}a _{i}, \label{CAction}
\end{eqnarray}
To compactify our notation we introduce $a _{i}^{b} (b=1,2)$
defined by $a _{i} =a _{i}^{1}, \; a _{i}^{\ast } =a _{i}^{2}.$ In
contrast to I, where we set  $V _{i}(t)=0$, here we allow the
presence of an external potential $V _{i}$ in $S[a _{i}^{\ast },a
_{i}]$. In the derivation of Boltzmann equations we will  assume
that $V _{i}$ is a slowly-varying function in position and time
and treat it as an external perturbation.

In terms of these fields the classical action takes the form

\begin{eqnarray}
S[a ]&=&\int dt\sum_{i}\frac{1}{2}h_{ab}a _{i}^{a}(t)\hbar\partial
_{t}a _{i}^{b}(t) \notag \\ + && \int dt\sum_{i}\left(J\sigma
_{ab}a _{i+1}^{a}(t)a_{i}^{b}(t) -\frac{1}{2}V_i(t)\sigma _{ab}a
_{i}^{a}(t)a_{i}^{b}(t)\right)\notag\\  &&-\int dt\sum_{i}\left(
\frac{U}{4 \mathcal{N}}(\sigma _{ab}a _{i}^{a}(t)a
_{i}^{b}(t))^{2}\right), \label{Caction}
\end{eqnarray}

\noindent where $\mathcal{N}$ is the number of fields, which is
two in this
case, and summation over repeated field indices $%
a,b=(1,2)$ is implied. $h_{ab}$ and $\sigma _{ab}$ are matrices
defined as

\begin{equation}
h_{ab}=i\left(
\begin{array}{cc}
0 & -1 \\
1 & 0
\end{array}
\right) \qquad \sigma _{ab}=\left(
\begin{array}{cc}
0 & 1 \\
1 & 0
\end{array}
\right)
\end{equation}
In terms of the familiar Pauli matrices, $\sigma_{ab}= \sigma_x$
and $ h_{ab}= - \sigma_y$.

We define the following index lowering convention
\begin{equation}
X_{a}=\sigma _{ab}X^{b}.
\end{equation}

After second quantization the fields $a _{i}^{a}$ are promoted to
operators. We denote the mean field or the expectation value of
the field operator by $z_{i}^{a}(t)$ (called $\phi_i^{a}(t)$ in
Paper I) and the expectation value of the fluctuation field
$\hat{\varphi}_{i}^{a}$,
 $\hat{\varphi}_{i}^{a}=\hat{a}_{i}^{a}-z_{i}^{a}$ by
$G_{ij}^{ab}(t,t^{\prime })$. Physically, $|z_{i}^{a}(t)|^{2}$ is
the condensate population and the two point functions
$G_{ij}^{ab}(t,t^{\prime })$ determines the quantum fluctuations
around the mean field:

\begin{eqnarray}
G_{ij}^{ab}(t,t^{\prime })  \equiv \left\langle
T_{C}{\hat{\varphi}} _{i}^{a}(t)\hat{\varphi} _{j}^{b}(t^{\prime
})\right\rangle
\end{eqnarray}

The brackets denote taking the expectation value with respect to
the density matrix and $T_{C}$ denotes time ordering along a
contour $C$ in the complex plane.

In order to describe the non-equilibrium dynamics we specify the
contour of integration  to be the
Schwinger-Keldysh contour \cite{ctp} along the real-time axis or \emph{%
closed time path } (CTP) contour. Using the CTP contour, the
two-point functions are decomposed as

\begin{equation}
G_{ij}^{ab}(t,t^{\prime })=\theta _{ctp}(t,t^{\prime
})G_{ij}^{ab>}(t,t^{\prime })+\theta _{ctp}(t^{\prime
},t)G_{ij}^{ab<}(t,t^{\prime }),
\end{equation}

\noindent where

\begin{eqnarray}
G_{ij}^{ab>}(t,t^{\prime }) &=&\left\langle \hat{\varphi}_{i}^{a}(t)%
\hat{\varphi}_{j}^{b}(t^{\prime })\right\rangle,\label{ggreat} \\
 G_{ij}^{ab<}(t,t^{\prime }) &=&\left\langle\hat{ \varphi}%
_{i}^{b}(t^{\prime
})\hat{\varphi}_{j}^{a}(t)\right\rangle,\label{gless}
\end{eqnarray}

\noindent with  $\theta _{ctp}(t-t^{\prime })$ being the CTP
complex contour ordered theta function defined  in Eq. (I.30).

All correlation functions of the quantum theory can be obtained
from the two particle irreducible (2PI) effective action $\Gamma
\lbrack z ,G]$. In Ref. \cite{byHFB} we showed  $\Gamma \lbrack z
,G]$ is given by:

\begin{eqnarray}
\Gamma \lbrack z ,G]&=&S[z ]+\frac{i\hbar}{2}Tr\ln  G^{-1}+\frac{i\hbar}{2}%
TrD^{-1}(z ) G\notag\\&& +\Gamma _{2}[z ,G]+const,  \label{2pi}
\end{eqnarray}
where $iD^{-1}(z )$ is the classical inverse propagator given by

\begin{eqnarray}
&&iD_{ijab}(t,t^{\prime })\;^{-1} =\frac{\delta S[z ]}{\delta z
_{i}^{a}(t)\delta z _{j}^{b}(t^{\prime })} \\
&&=\left( \delta _{ij}\hbar h_{ab}\partial _{t}+J(\delta
_{i+1j}+\delta
_{i-1j})\sigma _{ab}\right) \delta (t-t^{\prime })  \notag \\
&&-\frac{U}{\mathcal{N}}\left( 2z _{ia}(t)z _{ib}(t)+\sigma _{ab}z
_{i}^{c}(t)z _{ic}(t)\right) \delta _{ij}\delta (t-t^{\prime }),
\notag
\end{eqnarray}
and $\Gamma _{2}[z ,G]$ consists of all two-particle irreducible
vacuum graphs in the theory (the diagrams that do not become
disconnected by cutting two propagator lines) with propagators set
equal to $G$ and vertices determined by \ the \ interaction terms
in $S[ z +\varphi]$ .

The dynamical equations of motion for the mean field $z
_{i}^{a}(t)$ and the propagators $G_{ij}^{ab}(t,t^{\prime })$ are
found \ by solving the equations $ \frac{\delta \Gamma \lbrack z
,G]}{\delta z _{i}^{a}(t)} =0$ and $ \frac{\delta \Gamma \lbrack z
,G]}{\delta G_{ij}^{ab}(t,t^{\prime })}=0$. They were given in (I.
24) and (I.26) respectively.

The action \ $\Gamma $ including the full diagrammatic series for
$\Gamma _{2}$ gives the full dynamics. It is of course not
feasible to obtain an exact expression for $\Gamma _{2}$ in a
closed form. Various approximations for the full 2PI effective
action can be obtained by truncating the diagrammatic expansion \
for $\Gamma _{2}$. The ones relevant for this paper are the
\textbf{HFB approximation} and the \textbf{full second order
approximation}. The HFB approximation corresponds  to a truncation
of $\Gamma _{2}$ retaining only the first order diagram in $U$
which is $z$ independent, i.e. keeping only the \emph{double-
bubble} diagram (Fig. 1 in paper I). The full second order
approximation corresponds to a truncation retaining also diagrams
of second order in $U$ (
 the \emph{basket-ball} and  the \emph{setting-sun}).

Hereafter, to lighten the notation, we introduce a more compact
set of symbols for the physical quantities than was used in Paper
I, which contains more details:

\begin{eqnarray}
\textbf{z}(t_i)&=&\left(%
\begin{array}{c}
  \langle \hat{a}^1 _{i}(t)\rangle \\
   \langle \hat{a}^2 _{i}(t)\rangle  \\
\end{array}%
\right)=\left(%
\begin{array}{c}
  z(t_i) \\
   z^\ast(t_i) \\
\end{array}%
\right)\label{reddef1}\\
iH(t_i,t^{\prime }_j)&\equiv& z^{a}_i(t) {z _{j}}_{b}(t^{\prime }) \\
ig(t_i,t^{\prime }_j)&\equiv& {{G_{ij}}^{a}}_{b}(t,t^{\prime }),
\;\\
ig^{>}(t_i,t^{\prime }_j)&\equiv&
{{G_{ij}^{>}}^{a}}_{c}(t,t^{\prime }), \\
ig^{<}(t_i,t^{\prime}_j)&\equiv& {G_{ij}^{<}}^{a}}_{c}(t,t^{\prime
})= {{{G_{ji}^{>}}_{c}}^{a}(t^{\prime },t). \label{reddeff}
\end{eqnarray}
The notation $t_i$ means that the function must be evaluated at
the time $t$ and at the lattice site $i$.

\subsection{The HFB and full second order equations of motion}

The equations of motion derived from the 2PI-CTP effective action(
(I-24) and (I-26)) have terms which can be grouped as the single
particle, the HFB and the second order contributions, as follows:


\begin{widetext}
\begin{eqnarray}
\sum_{k} \int dt^{\prime \prime }\left( D_{o}^{-1}(t_i,t_k^{\prime
\prime })-S^{HFB}(t_i,t_k^{\prime \prime })\right) H(t_k^{\prime
\prime },t_j^\prime) &=& \sum_{k} \int dt^{\prime \prime
}{{S}}(t_i,t_k^{\prime \prime })
H(t_k^{\prime \prime },t_j^{\prime }),  \label{condkb} \\
\sum_{k} \int dt^{\prime \prime }\left( {{D_{o}^{-1}}}
(t_i,t_k^{\prime \prime })-{{\Sigma ^{HFB}}}(t_i,t_k^{\prime
\prime })\right) g(t_k^{\prime \prime },t_j^{\prime  }) &=&
\sum_{k} \int dt^{\prime \prime }\Sigma (t_i,t_k^{\prime \prime })
g(t_k^{\prime \prime },t_j^{\prime  })-\delta _{ij}\delta
_{C}(t-t^{\prime }), \label{fluckb}
\end{eqnarray}
\end{widetext}
 where
${D_{o}^{-1}}(t_i,t_j^{\prime })$ is the inverse free particle
propagator given by:
\begin{eqnarray}
{D_{o}^{-1}}(t_i,t_j^{\prime }) &\equiv &\left( i\delta _{ij}
\sigma_z\partial _{t}+J(\delta _{i+1j}+\delta _{i-1j}) \right)
\delta (t-t^{\prime })\notag\\&&-\delta _{ij}V _{i} \delta
(t-t^{\prime }),
\end{eqnarray}
\noindent $\sigma_z$ is  the Pauli matrix:

\begin{eqnarray}
\sigma_z&=&\left(\begin{array}{cc}
  1 & 0 \\
  0 & -1
\end{array}\right),
\end{eqnarray}

\noindent  $S^{HFB}$ and $\Sigma ^{HFB}$  are the HFB
self-energies of $H$ and $g$ respectively, and $S$ and $\Sigma$
are the remaining parts of the self-energies of $H$ and $g$, which
we will assume are given by the second order corrections.

Using Eq. (I-43) it can be shown that $S^{HFB}$ and $\Sigma
^{HFB}$ are given by:

\begin{widetext}

\begin{eqnarray}
\Sigma ^{HFB}(t_i,t_j^{\prime }) &\equiv&
i\frac{U}{\mathcal{N}}\left( Tr\left (H(t_i,t_j^{\prime
})+g(t_i,t_j^{\prime })\right)I+2\left(H(t_i,t_j^{\prime
})+g(t_i,t_j^{\prime })\right) \right) \delta (t-t^{\prime
})\delta_{ij,} ,  \label{hfbGkb}    \\
{S^{HFB}(t_i,t_j^{\prime }) } &\equiv
&i\frac{U}{\mathcal{N}}\left( Tr\left (H(t_i,t_j^{\prime })
+g(t_i,t_j^{\prime })\right)I+ 2 g(t_i,t_j^{\prime }) \right)
\delta (t-t^{\prime })\delta _{ij}. \label{hfbpkb}
\end{eqnarray}

\end{widetext}
\noindent where $I$ is the identity matrix.

In paper I, we used the CTP contour of integration (which is also
usually called "in-in" contour) to evaluate \textbf{the second order
contribution}. Use of the CTP formalism was important there, because
it provided the technical means to formulate our initial value
problem in a completely causal manner, removing the Feynman boundary
conditions on the Green's function used in the conventional "in-out"
formalism \cite{Kadanoff}. In this work we are more interested in
deriving kinetic equations which are especially devised to study
relaxation of systems close to equilibrium. With this purpose in
mind, as done  by Kadanoff and Baym \cite{Kadanoff,Kane}, it is
better to set the initial conditions in the far past. We follow them
hereafter and  use the CTP contour, but instead of setting the
initial time to zero, as was done in paper I, we choose it to be
$-\infty$. The equations of motion we obtain in this way agree with
the equations of motion \cite{Kadanoff,Kane} and are given by:

\begin{widetext}
\begin{eqnarray}
&& \sum_{k} \int_{-\infty }^{\infty }dt^{\prime \prime }\left(
D_{o}^{-1}(t_i,t_k^{\prime \prime }) -S^{HFB}(t_i,t_k^{\prime
\prime })\right)H(t_k^{\prime \prime},t_j^{\prime}) -
\int_{-\infty }^{t}dt^{\prime \prime }
\gamma(t_i,t_k^{\prime\prime })H(t_k^{\prime \prime
},t_j^{\prime})=0,  \label{meankb} \\
&&\sum_{k} \int_{-\infty }^{\infty }dt^{\prime \prime
}H(t_i,t_k^{\prime\prime}) \left( D_{o}^{-1}(t_k^{\prime \prime}
,t_j^{\prime}) -S^{HFB}(t_k^{\prime \prime },t_j^{\prime})\right)
+\int_{-\infty }^{t^\prime }dt^{\prime \prime
}H(t_i,t_k^{\prime\prime}) \gamma(t_k^{\prime\prime
},t_j^{\prime})=0,\label{meankb2}
\end{eqnarray}
\begin{eqnarray}
&& \sum_{k} \int_{-\infty }^{\infty }dt^{\prime \prime }\left(
{D_{o}^{-1}(t_i,t_k^{\prime \prime })}-{\Sigma
^{HFB}}(t_i,t_k^{\prime \prime })\right) g^{(\gtrless
)}(t_k^{\prime \prime },t_j^{\prime })  = \label{former} \\&&
\sum_{k} \int_{-\infty }^{t}dt^{\prime \prime }\Gamma
(t_i,t_k^{\prime \prime }) g^{(\gtrless )}(t_k^{\prime \prime
},t_j^{\prime }) -\sum_{k} \int_{-\infty }^{t^{\prime }}dt^{\prime
\prime }\Sigma^{(\gtrless )}(t_i,t_k^{\prime \prime
})A(t_k^{\prime \prime },t_j^{\prime
})\notag, \\
&& \sum_{k} \int_{-\infty }^{\infty }dt^{\prime \prime
}g^{(\gtrless )}(t_i,t_k^{\prime \prime })\left(
{D_{o}^{-1}}(t_k^{\prime \prime },t_j^{\prime })-{\Sigma
^{HFB}}(t_k^{\prime \prime },t_j^{\prime })\right) = \label{last}\\
&& \sum_{k} \int_{-\infty }^{t}dt^{\prime \prime
}A(t_i,t_k^{\prime \prime })\Sigma ^{(\gtrless )}(t_k^{\prime
\prime },t_j^{\prime })\notag
  -\sum_{k} \int_{-\infty }^{t^{\prime }}dt^{\prime
\prime }{g^{(\gtrless )}}(t_i,t_k^{\prime \prime
})\Gamma(t_k^{\prime \prime },t_j^{\prime }).\notag
\end{eqnarray}

\end{widetext}
\noindent In the above equations,  Eq. (\ref{last})
is the hermitian conjugate of Eq. (\ref{former}), Eq.
(\ref{meankb2}) is the hermitian conjugate of Eq. (\ref{meankb})
and we have introduced the spectral functions
\begin{eqnarray}
\gamma(t_i,t_j^{\prime}) &\equiv
&(S^{>}(t_i,t_j^{\prime})-S^{<}(t_i,t_j^{\prime})), \\
\Gamma (t_i,t_j^{\prime \prime }) &\equiv
&(\Sigma^{>}(t_i,t_j^{\prime \prime })-\Sigma^{<}(t_i,t_j^{\prime
\prime
})), \\
A(t_i,t_j^{\prime \prime }) &\equiv &(g^{>}(t_i,t_j^{\prime \prime
})-g^{<}(t_i,t_j^{\prime \prime })).
\end{eqnarray}

\noindent Notice that $A(t_i,t_j^{\prime \prime
}),\Gamma(t_i,t_j^{\prime \prime }),\gamma(t_i,t_j^{\prime \prime
})$ are just the
spectral functions defined in (I. 36) 
multiplied by a minus sign. In Paper I we denoted these by a
subscript ${}^{(\rho)}$. Here, for ease of comparisons with the
literature, we have changed to the notation of Kadanoff and Baym
\cite{Kadanoff}, $\gamma$, $\Gamma$ and $A$. We will show later
that $\gamma$ and $\Gamma$ contain information about the
condensate and noncondensate particle decay rates respectively.

If we use the full second order expansion,  (I.60) 
and (I.61), 
$S^{(\gtrless )}$ and $\Sigma^{(\gtrless )}$ are given by
\begin{widetext}

\begin{eqnarray}
S^{(\gtrless )}(t_i,t_j^{\prime })&\equiv& -\frac{1}{2}\left(
\frac{2U}{\mathcal{N}} \right) ^{2}\left( g^{(\gtrless
)}(t_i,t_j^{\prime })    Tr\left(g^{(\gtrless )}(t_i,t_j^{\prime
})g^{(\lessgtr )}(t_j^{\prime},t_i)\right)\right.  + \left.2
g^{(\gtrless )}(t_i,t_j^{\prime })
g^{(\lessgtr)}(t_j^{\prime},t_i)g^{(\gtrless )}(t_i,t_j^{\prime
})\right),
\end{eqnarray}

\begin{eqnarray}
&&\Sigma^{(\gtrless )}(t_i,t_j^{\prime })\equiv -\frac{1}{2}\left(
\frac{2U}{\mathcal{N}}\right) ^{2}\times \left\{ H(t_i,t_j^{\prime
}) Tr\left(g^{(\gtrless )}(t_i,t_j^{\prime })g_{ik}^{(\lessgtr
)}(t_j^{\prime},t_i)\right)+2 H(t_i,t_j^{\prime }) g^{(\lessgtr
)}(t_j^{\prime},t_i)g^{(\gtrless )}(t_i,t_j^{\prime })\right. \notag \\
&&+2g^{(\gtrless )}(t_i,t_j^{\prime })\left( H(t_j^{\prime},t_i)
g^{(\gtrless)}(t_i,t_j^{\prime })+ g^{(\lessgtr
)}(t_j^{\prime},t_i)H(t_i,t_j^{\prime }) +g^{(\lessgtr
)}(t_j^{\prime},t_i)g^{(\gtrless )}(t_i,t_j^{\prime })\right)   \\
&&\left. g^{(\gtrless )}(t_i,t_j^{\prime })Tr\left(
H(t_j^{\prime},t_i) g^{(\gtrless )}(t_i,t_j^{\prime})+g^{(\lessgtr
)}(t_j^{\prime},t_i)H(t_i,t_j^{\prime })+g^{(\lessgtr
)}(t_j^{\prime},t_i)g^{(\gtrless )}(t_i,t_j^{\prime })\right)
\right\}. \notag
\end{eqnarray}
\end{widetext}

It is convenient to decompose the above equations in their matrix
components. To do that we introduce the definitions

\begin{eqnarray}g^{>}(t_i,t_j^{\prime })&=&-i\left(\begin{array}{cc}
  \widetilde{\rho}_{ij} (t,t^{\prime })& m_{ij}(t,t^{\prime }) \\
  m^\ast_{ji}(t^{\prime },t) & \rho_{ji}(t^{\prime },t)
\end{array}\right), \label{defmat1}\\
g^{<}(t_i,t_j^{\prime })&=&-i \left(\begin{array}{cc}
  \rho_{ij}(t,t^{\prime }) & m_{ji}(t^{\prime },t) \\
  m^\ast_{ij}(t,t^{\prime }) & \widetilde{\rho}_{ji}(t^{\prime },t)
\end{array}\right). \label{defmat2}
\end{eqnarray}
\noindent At equal times, the quantities $\widetilde{\rho}_{ij} $
and  $\rho_{ij} $  are related by the bosonic commutation
relations. Using Eqs. (\ref{defmat1}) and (\ref{defmat2}) into the
self-energy equations we get

\begin{eqnarray}
S^{>}_{11}(t_i,t_j^{\prime })&=& \frac{ -i8U^{2}}{\mathcal{N}^{2}}
  \widetilde{\rho}_{ij}(2 m_{ij} m^*_{ji}+\widetilde{\rho}_{ij}
  \rho_{ji}),\\
S^{>}_{12}(t_i,t_j^{\prime })&=& \frac{ -i8U^{2}}{\mathcal{N}^{2}}
   m_{ij}( m_{ij} m^*_{ji}+ 2 \widetilde{\rho}_{ij} \rho_{ji}),\\
S^{<}_{11}(t_i,t_j^{\prime })&=&   \frac{
-i8U^{2}}{\mathcal{N}^{2}}
  \rho_{ij}(2 m_{ji} m^*_{ij}+\widetilde{\rho}_{ji} \rho_{ij}),\\
S^{<}_{12}(t_i,t_j^{\prime })&=&   \frac{
-i8U^{2}}{\mathcal{N}^{2}} m_{ji}( m_{ij} m^*_{ji}+ 2
\widetilde{\rho}_{ji} \rho_{ij} ),
\end{eqnarray}

\begin{widetext}

\begin{eqnarray}
\Sigma ^{HFB} (t_i,t_j^{\prime })&=&\frac{2U}{\mathcal{N}}\left(
\begin{array}{cc}
2|z_i|^2+ \rho_{ii}+\widetilde{\rho}_{ii}& z_i^{2}
+m_{ii}\\
 z_i^{*2} +m_{ii}^* & 2|z_i|^2+
\rho_{ii}+\widetilde{\rho}_{ii}
\end{array}
\right) \delta{(t-t^{\prime })}\delta_{ij}, \\
S^{HFB}(t_i,t_j^{\prime }) &=&\frac{2U}{\mathcal{N}} \left(
\begin{array}{cc}
|z_i|^2+\rho_{ii}+\widetilde{\rho}_{ii} & m_{ii} \\
m_{ii}^* &  |z_i|^2+\rho_{ii}+\widetilde{\rho}_{ii}
\end{array}
\right)\delta{(t-t^{\prime })}\delta_{ij},
\end{eqnarray}

\begin{eqnarray}
 \Sigma^{>}_{11}(t_i,t_j^{\prime })&=&
\frac{ -i8U^{2}}{\mathcal{N}^{2}}(
\rho_{ji}{{\widetilde{\rho}_{{ij}}}}^2 +
2{m_{{ij}}}{\widetilde{\rho}_{{ij}}}{{{m_{{ji}}}}^*}
+2{\widetilde{\rho}_{{ij}}}{{{m_{{ji}}}}^*}{z_i}{z_j} +
{{\widetilde{\rho}_{{ij}}}}^2{z_j}{{{z_i}}^*} +
  2\rho_{ji}{\widetilde{\rho}_{{ij}}}{z_i}{{{z_j}}^*} + \\&& \notag
  2{m_{{ij}}}{{{m_{{ji}}}}^*}{{{z_j}}^*}{z_i} +
  2{m_{{ij}}}{\widetilde{\rho}_{{ij}}}{{{z_i}}^*}{{{z_j}}^*}),\\
\Sigma^{>}_{12}(t_i,t_j^{\prime })&=&  \frac{
-i8U^{2}}{\mathcal{N}^{2}}(2\rho_{ji}{m_{{ij}}}{\widetilde{\rho}_{{ij}}}
+
  2\rho_{ji}{\widetilde{\rho}_{{ij}}}{z_i}{z_j} +
{{m_{{ij}}}}^2{{{m_{{ji}}}}^*}
  +
  2{m_{{ij}}}{z_i}{z_j}{{{m_{{ji}}}}^*} +
  2{m_{{ij}}}{\widetilde{\rho}_{{ij}}}{z_j}{{{z_i}}^*} + \\\notag&&
  2\rho_{ji}{m_{{ij}}}{z_i}{{{z_j}}^*} +
  {{m_{{ij}}}}^2{{{z_i}}^*}{{{z_j}}^*}),\\
\Sigma^{<}_{11}(t_i,t_j^{\prime })&=&  \frac{
-i8U^{2}}{\mathcal{N}^{2}}({\rho_{ij}}^2{\widetilde{\rho}_{ji}} +
2\rho_{ij}{m_{{ji}}}{{{m_{{ij}}}}^*} +
  2{\rho_{ij}}{z_i}{z_j}{{{m_{{ij}}}}^*} + \rho_{ij}^2{z_j}{{{z_i}}^*}
  +
  2{\rho_{ij}}{\widetilde{\rho}_{{ji}}}{z_i}{{{z_j}}^*} + \\&& \notag
  2{m_{{ji}}}{z_i}{{{m_{{ij}}}}^*}{{{z_j}}^*} +
  2{\rho_{ij}}{m_{{ji}}}{{{z_i}}^*}{{{z_j}}^*}),\\
\Sigma^{<}_{12}(t_i,t_j^{\prime })&=&  \frac{
-i8U^{2}}{\mathcal{N}^{2}}(2{\rho_{ij}}{m_{{ji}}}{\widetilde{\rho}_{{ji}}}
+ 2{\rho_{ij}}{\widetilde{\rho}_{{ji}}}{z_i}{z_j} +
{{m_{{ji}}}}^2{{{m_{{ij}}}}^*}
  +
  2{m_{{ji}}}{z_i}{z_j}{{{m_{{ij}}}}^*} +
  2{\rho_{ij}}{m_{{ji}}}{z_j}{{{z_i}}^*} + \\&&
  2{m_{{ji}}}{\widetilde{\rho}_{{ji}}}{z_i}{{{z_j}}^*} +
  {{m_{{ji}}}}^2{{{z_i}}^*}{{{z_j}}^*})\notag,
\end{eqnarray}

\end{widetext}
and
\begin{eqnarray}
 S^{\gtrless}_{22}(t_i,t_j^{\prime
})&=&S^{\gtrless}_{11}(t_i,t_j^{\prime
})\left\{\rho_{{ij}}\rightleftarrows{{{\widetilde{\rho}_{{ij}}}}}\right\},\\
S^{\gtrless}_{21}(t_i,t_j^{\prime
})&=&S^{\gtrless}_{12}(t_i,t_j^{\prime
})\left\{m_{{ji}}\rightleftarrows{{{m_{{ij}}}}^*}\right\},\\
\Sigma^{\gtrless}_{22}(t_i,t_j^{\prime
})&=&\Sigma^{\gtrless}_{11}(t_i,t_j^{\prime
})\left\{{z_i}\rightleftarrows{z_j},\rho_{{ij}}\rightleftarrows{{{\widetilde
{\rho}_{{ij}}}}}\right\},\\
\Sigma^{\gtrless}_{21}(t_i,t_j^{\prime
})&=&\Sigma^{\gtrless}_{12}(t_i,t_j^{\prime
})\left\{{z_i}\rightleftarrows{z_j}^*,m_{{ji}}\rightleftarrows{{{m_{{ij}}}}^
*}\right\}.
\end{eqnarray}

\noindent The above expressions for the self-energy, which contain
two-particle irreducible diagrams up to second order in the
interaction strength, agree exactly with those used in
Refs.\cite{Millena,Wachter,Holland2}. In Ref.
\cite{Wachter,Holland2}  the authors used these equations as the
starting point of a quantum kinetic theory before applying the
Markovian approximation. It is important to mention that in
contrast to other self-energy approximations that may lead to
equations of motion that do not satisfy conservation laws, the 2PI
effective action formalism is a "$\Phi$-derivable" \cite{Martin,
Baym} approximation and therefore all the equations of motion
derived from it are guaranteed to be conserving. Moreover, as we
showed in paper I, a truncation up to second order in the
interaction strength is not appropriate to describe
far-from-equilibrium dynamics  outside the weak coupling regime.
Away from the weak coupling regime, the 2PI effective action can
be a powerful tool. For example a $1/\mathcal{N}$ expansion of the
2PI effective action has been shown to provide a practicable
controlled nonperturbative description of far-from equilibrium
dynamics without the small coupling restriction\cite{Berges}.

\section{From quantum theory of interacting particles to quantum kinetic
theory} \label{sec3}

From previous sections it can be observed that the equations of
motion obtained from the 2PI effective action are quite involved:
nonlinear and nonlocal integro-differential equations, not readily
solvable in closed form. To progress  further we need to introduce
approximations based on physical considerations.  It is easier to
proceed if one can observe and justify a separation of time scales
in the relevant physical processes in question, i.e., one related
to quantum processes which are usually microscopic in scale (note
quantum entanglement and correlation of the system can extend to
much greater scales, meso or even macro) and one related to the
kinetic or transport properties, which is usually macroscopic in
scale. However, this assumption of a scale separation, may not be
valid in mesoscopic processes (as in strongly correlated systems)
or macroscopic quantum coherence effects (see footnote 1). For
those situations where a separation of macroscopic and microscopic
time scales which would permit an effective kinetic theory
description does {\em not\/} exist, one can adopt the
\textit{effectively open system} framework quantified by the
\textit{n}PI-CTP effective action and the hierarchy of equations
it generates.  We begin with a discussion of the latter situation
which is more demanding and general. We describe the conceptual
pathway for the construction of quantum kinetic theory from the
\textit{n}PI effective action. Though somewhat theoretical and
formally oriented, it may be of some use, as this is the first
point of contact with quantum kinetic theory from the effective
action approach, in the atomic and molecular physics (AMO)
context(For more details on this subject  see
\cite{RamseyPhD,CH88,mea,CH03}, where our discussions in the
following section are based on).

\subsection{Quantum kinetic theory from (\textit{n}PI) effective action}

It may be useful to begin by defining what we mean by a quantum
kinetic theory. It contains,  but supersedes,  the quantum version
of Boltzmann's theory.  Formally it refers to the theory based on
the hierarchy of coupled equations for the (relativistic) Wigner
function\cite{Wigner}  and its higher-correlation analogs, which
are obtained by a Fourier transform of the relative coordinates in
the Schwinger-Dyson equations\cite{FetterW}  for the correlation
functions, or alternatively, in the \emph{master effective action}
(defined as the \textit{n}PI effective action when $n \rightarrow
\infty$, we are dealing with $n=2$ here) whose variation yields
the Schwinger-Dyson equations. This is a quantum analogue of the
BBGKY hierarchy \cite{Balescu}, expressed in a representation
convenient for distinguishing between microscopic (quantum
field-theoretic) and macroscopic (transport and relaxation)
phenomena.  As such, it does not require near-equilibrium
conditions, and in fact, is applicable for a rather general moment
expansion of the initial density matrix \cite{CH88}. \footnote{It
should be pointed out that in order to {\em identify\/} the Wigner
function with a distribution function for quasiparticles, one must
show that the density matrix has {\em decohered,\/} and this is
neither guaranteed nor required by the existence of a separation
of macroscopic and microscopic time scales \cite{habib:1990a}.}

To understand how quantum kinetic theory is derived from an
\textit{n}PI effective action and how it relates to the familiar
Boltzmann's theory, it is perhaps helpful to examine the relation
between this theory in its full generality and an effective
Boltzmann description of relaxation phenomena for the one-particle
distribution function of quasiparticles.  In nonequilibrium
statistical mechanics, as is well known
\cite{Balescu,spohn:1983a}, the act of truncating the classical
BBGKY hierarchy does not in itself lead to irreversibility and an
$H$-theorem.  One must further perform a type of {\em coarse
graining\/} of the truncated, coupled equations for $n$-particle
distribution functions.  For example, if one truncates the
hierarchy to include only the one-particle and two-particle
distribution functions, it is the subsequent assumption that the
two-particle distribution function at some initial time
{factorizes} in terms of a product of single-particle distribution
functions (which is at the heart of the molecular chaos hypothesis
where the colliding particles are initially independent, but
correlated after a collision ) what leads to the (irreversible)
Boltzmann equation. The assumption that the two-particle
distribution function factorizes is an example of a type of coarse
graining called {\em slaving\/} of the two-particle distribution
function to the single-particle distribution function, in the
language of \cite{mea}. The situation in quantum kinetic field
theory is completely analogous. One may choose to work with a
truncation of the hierarchy of the Wigner function and its higher
correlation analogs, or one may instead perform a slaving of, for
example, the Wigner-transformed four-point function, which leads
(within the context of perturbation theory) directly to the
(relativistic) Boltzmann equation \cite{CH88} and the usual
$H$-theorem \cite{CH03}. Typically this slaving of the higher
correlation function(s) involves imposing causal boundary
conditions to obtain a particular solution for the higher
correlation function(s) in terms of the lower order correlation
functions \cite{CH88,mea}. The truncation and subsequent slaving
of the hierarchy within quantum kinetic field theory can be
carried out at any desired order, as dictated by the initial
conditions and relevant interactions. As with any coarse graining
procedure, in implementing the slaving of a higher
correlation/distribution function to lower
correlation/distribution functions, one is going over from a
closed system to an {\em effectively open system,\/} the hallmarks
of which are the emergence of dissipation \cite{CH88} and
noise/fluctuations \cite{mea}. This fact has led some to search
for stochastic generalizations of the Boltzmann equation
\cite{KacLog}, motivated by the fact that systems in thermal
equilibrium always manifest fluctuations, as embodied in the
fluctuation-dissipation relation. (A derivation of the stochastic
Boltzmann equation from quantum field theory can be found in
\cite{mea}.)

The essential point about the process of slaving of higher
correlation (or distribution) functions is that it is a step which
is  {\em independent\/} of the assumption of macroscopic and
microscopic time scales.  In fact, a completely analogous
procedure exists at the level of the Schwinger-Dyson equations
(i.e., without Wigner transformation) for correlation functions in
an interacting quantum field theory \cite{mea}. Recall that the
Schwinger-Dyson equations are, in the context of nonequilibrium
field theory in the Schwinger-Keldysh or closed-time-path (CTP)
formulation, an infinite chain of coupled dynamical equations for
all order correlation functions of the quantum field.  The
importance of the closed-time-path formalism in nonequilibrium
situations is that it ensures that the equations are causal and
that the correlation functions are ``in-in'' expectation values in
the appropriate initial quantum state or density matrix.  As with
the BBGKY hierarchy in nonequilibrium statistical mechanics, the
common strategy is  to truncate the hierarchy of correlation
functions at some finite order. A general procedure has been
presented for obtaining coupled equations for correlation
functions at any order $l$ in the correlation hierarchy, which
involves a truncation of the {master effective action\/} at a
finite order in the loop expansion \cite{mea}.  By working with an
$l$ loop-order truncation of the master effective action, one
obtains a closed, time-reversal invariant set of coupled equations
for the first $l+1$ correlation functions, $z=C_1$, $g=C_2$,
$C_3$, \ldots, $C_{l+1}$. In general, the equation of motion for
the highest order correlation function will be linear, and thus
can be formally solved using Green's function methods.  The
existence of a unique solution depends on supplying causal
boundary conditions. When the resulting solution for the highest
correlation function is then back-substituted into the evolution
equations for the other lower-order correlation functions, the
resulting dynamics becomes non- time-reversal invariant, and
generically dissipative. As with the slaving of the higher-order
Wigner-transformed correlation functions in quantum kinetic field
theory, we have then gone over from a closed system (the truncated
equations for correlation functions) to an {effectively open
system.\/}  In addition to dissipation, one expects that an
effectively open system will manifest noise/fluctuations (an
example of slaving the four-point function to the two-point
function in the symmetry-unbroken $\lambda \Phi^4$ field theory is
given in \cite{mea}).  Thus a framework exists for exploring
irreversibility and fluctuations within the context of a unitarily
evolving quantum field theory, using the truncation and slaving of
the correlation hierarchy.

While it is certainly not the only coarse-graining scheme which
could be applied to an interacting quantum field, the slaving of
higher correlation functions to lower-order correlation functions
within a particular truncation of the correlation hierarchy, as a
particular coarse graining method, has several important benefits.
First, it can be implemented in a truly nonperturbative fashion,
where the variance of the mean field can be on the order of the
"classical" mass (defined as the second order derivative of the
effective potential in the equation of motion for the mean field,
which provides the natural time scale of the system dynamics).
This necessitates a nonperturbative resummation of daisy graphs
(the leading contributions in a large $\mathcal{N}$ expansion)
\cite{2pi}, which can be incorporated in the truncation/slaving of
the correlation hierarchy in a natural way. \footnote{At late
times in the thermalization stage, when the quantum field is near
equilibrium, an effective kinetic description may be justified,
but will likely require resummation of  hard thermal loops (see,
e.g., \cite{BlaIan}) . Under such circumstances, even the
evaluation of transport coefficients is nontrivial for high
temperatures \cite{jeon:1995a,CHR}.} Second, the truncation of the
correlation hierarchy accords with our intuition that the degrees
of freedom readily accessible to physical measurements are often
limited to the mean field and two-point function.

\section{Systems whose dynamics admit two-time separation}
\label{Boltzmann}

An alternative (actually more common and easier) route to reach a
kinetic theory description from n-body quantum dynamics becomes
available  when there is a clear separation of two time scales in
the system dynamics. This is the usual text book treatment of
kinetic theory we are familiar with. The two different scales in
the system  are the time (or length) scale separation between the
duration of a collision event (or scattering length) and the
inverse collision rate (or the mean free path). For quantum
processes,  in the weakly interacting regime, we expect there is
also a separation between the kinetic scale of n particles
(expressed in the center of mass coordinate) and the quantum scale
(expressed in the momentum corresponding to the Fourier transform
of the relative coordinates between two particles), which
describes how quantum processes (such as radiative corrections)
change the particles' mass-energy and momenta.  Using these
approximations it is possible to recast the full quantum dynamics
into the simpler forms of two coupled equations which constitute
quantum kinetic theory, the Boltzmann equation governing the
distribution functions and what is known as the gap equation for
the modified dispersion relation.

\bigskip

 For a three dimensional uniform Bose gas the
duration of a collision event $\tau_0$ is  given by the time that a
particle with average velocity $v$ spends in the interaction region
measured by the range of the two-particle interaction potential.
This range for a repulsive potential  is  typically given by the
$s$-wave scattering length and thus  $\tau_0\approx a_s/v$. On the
other hand, the inverse collision rate $\tau_c$  or  time between
successive collisions  is approximately given  by $\tau_c\approx(n
a_s^2 v)^{-1}$, where $n$ is the particle density. The required
separation of time scales, $\tau_c\gg\tau_0$ implies the inequality
$na_s^3\ll1$ or in other words, the necessary condition required for
the validity of a scale separation is that the system must be in the
dilute weakly interacting regime.
For atoms in optical lattices the
dilute weakly interacting conditions required for the scale
separation is fulfilled if the average repulsive interaction energy
 $Un$, where n is the mean number of particles per lattice
site, is much smaller than $J$, the quantum kinetic energy needed to
correlate two atoms at adjacent lattice sites, or $Un/J\ll1$.

\bigskip
Perhaps an intuitive way to understand the scale separation is the
following. At equilibrium the correlation functions describing a
homogeneous system are translationaly invariant and stationary. If
the system is disturbed from equilibrium,  collisions among
particles would break both invariances. However,  as long as the the
interaction energy per particle is smaller than the  typical kinetic
energy per particle, inter-particle collisions are few and far
between. In this case the  quantum-mechanical entanglement between
collision partners decays faster than the time required for the next
collision to take place, particles can be considered as free between
collisions and approximate time and space translational invariance
holds.

\subsection{Coarse-graining procedure}

To make the scale separation, for BEC systems at hand, it is best
to perform first a gauge transformation which makes it easier to
identify (and coarse-grain away)  the \ fast variations induced by
the rapid change of the condensate phase. Following Ref.
\cite{Kane} we introduce the gauge transformation
\begin{eqnarray}
z(t_i) &=&e^{ i \theta (t_{i})}\sqrt{n_o(t_i)}, \\
g^{(\gtrless )}(t_i,t_j^{\prime }) &=&e^{ i \theta
(t_{i})\sigma_z}\tilde{g}^{(\gtrless )}(t_i,t_j^{\prime })e^{ -i
\theta (t^{\prime }_{j})\sigma_z},
\end{eqnarray}

\noindent where $\sqrt{n_{o}(t_i)}$ and $\theta (t_{i})$ are real.
The equations of motion are invariant under the phase
transformation if we replace $D_{o}^{-1}$ by $\tilde{D}_{o}^{-1}$:
\begin{widetext}
\begin{eqnarray}
\tilde{D}_{o}^{-1}(t_i,t_j^{\prime })=\left( \hbar \delta
_{ij}(i\sigma_z\partial _{t}-\partial _{t}\theta(t _{i}))-\delta
_{ij}V _{i}+J(e^{i\sigma_z \Delta \theta (t _{i+1/2})}\delta
_{i+1j}+e^{-i\sigma_z \Delta \theta (t _{i-1/2})}\delta
_{i-1j})\right) \delta (t-t^{\prime }),
\end{eqnarray}
\end{widetext}
 \noindent where we have introduced the definition
$\Delta \theta(t _{i+1/2})= \theta(t _{i+1})-\theta (t_{i})$. As
shown in Ref. \cite{Bogoliubov}, in the context of the discrete
Bose-Hubbard model, it is convenient to map the unitary gauge
transformation to the so called phase-twist of the Hamiltonian.
The twisted Hamiltonian exhibits additional phase factors, $e^{\pm
i \Delta \theta }$ in the hopping term, which are known as the
Peierls phase factors \cite{Poilblanc,ShastrySutherland}.

The scale separation is performed by introducing  the variables:

\begin{equation}
R=(i+j)/2, \quad T=(t+t^{\prime })/2,
\end{equation}

\begin{equation}
r=(i-j), \quad \tau =(t-t^{\prime }),
\end{equation}

\noindent  For a translationally invariant system at equilibrium,
 the condensate density $n_o(t_i)$  is  position and time
independent and the propagators $g^{(\gtrless )}(t_i,t_j^{\prime })$
only depend on the relative coordinates variables $r$ and $\tau$ and
are highly peaked about their zeros. If  the system is disturbed by
small perturbations, such as an external potential $V(t_{i})$ which
varies slowly in space and time, we expect for systems with scale
separation, that  the gauge-transformed propagators,
$\tilde{g}^{(\gtrless )}(t_i,t_j^{\prime })$,   acquire  a slowly
varying dependence on the center of mass coordinates $R$ and $T$ but
still to be peaked around the zeros of $r$ and $\tau$. We emphasis
that the gauge transformed, not the original variables, are the ones
that are expected to be slowly varying. The reason is that even if
the perturbation is slowly varying, the phase $\theta(t_i)$ can be a
rapidly varying function and it can induce strong variations in the
condensate amplitude and in the propagators.

Before going further, it is important to discuss the issue that by
defining the spatial center of mass coordinates at points that
strictly speaking are not lattice sites points we might be
introducing un-physical degrees of freedom. We stress though  that
this is not the case for system with scale separation. Under the
slowly varying approximation the un-physical degrees of freedom are
excluded, since the functions evaluated at the $R$ points may be
thought of as the average over neighboring physical lattice sites.

\bigskip

We proceed now to describe the coarse-graining procedure that uses
the slowly varying property of the propagators in the center of mass
variables to simplify the equations of motion.

 If the phase twist applied to
the system is small $\Delta \theta\ll\pi$, the Peierls phase
factors can be written as, $e^{\rm{i}\Delta \theta}=1-i\Delta
\theta- \frac{1}{2}\Delta \theta^2$.  In this case, the phase
factors can be physically connected  to the imposition of an
acceleration on the lattice and the energy change resulting from
the phase twist can be attributed to the kinetic energy of the
superflow generated by the acceleration. Under this picture in the
context of the Bose-Hubbard model the quantity $\Delta \theta$ can
be also connected, as is the gradient of the phase in non lattice
systems, to the superfluid velocity:
\begin{eqnarray}
\hbar {v_{s}}(t_{i+1/2})&=&2J\Delta \theta (t_{i+1/2})a_l.
\label{suv}
\end{eqnarray}
with $a_l$ the lattice spacing.

If the disturbances introduced by the perturbation are small, the
superfluid velocity is expected  to be a slowly varying function
in space and time and to a good approximation  its second order
variations can be ignored, i.e. $\Delta v_{s}(t)~\equiv
~2[v_{s}(t_{i+1/2})-v_{s}(t_{i})], \forall \quad i$. Again, the
quantity $v_{s}(t_{i}))$  may be thought of as the average over
neighboring lattice sites:
$[v_{s}(t_{i+1/2})+v_{s}(t_{i-1/2})]/2$. Using the small angle and
slowly varying dependence of the superfluid velocity,  the
propagator $\tilde{D}_{o}^{-1}$ can thus be written in terms of
the superfluid velocity  as:
\begin{widetext}
\begin{eqnarray}
\tilde{D}_{o}^{-1}(t_i,t_j^{\prime })&\approx& \left(\delta _{ij}
[{\rm i} \hbar
 \sigma_z \partial_t-\hbar\partial_t \theta(t
_{i})-V (t_{i})-J\overline{v}_{s}^2(t_i)]+ J(1+
\frac{\rm{i}}{2}\sigma_z \Delta \overline{v}_{s}(t)) [\delta
_{i+1j}+\delta _{i-1j}]  \right)\delta (t-t^{\prime})\notag \\
&+&\left(J {\rm{i}} \sigma_z \overline{v}_{s}(t_i)( \delta
_{i+1j}-\delta_{i-1j}) \right)\delta (t-t^{\prime}).
\end{eqnarray}
\end{widetext}

\noindent where we have introduced the dimensionless superfluid
velocity ${\overline{v}_{s}}(t_i)\equiv \frac{\hbar {v_{s}}(t_i)}{2J
a_l}$.

At equilibrium, the time derivative of the phase is related  to the
chemical potential. Extending this identification  to the
nonequilibrium system we define the chemical potential as
\begin{eqnarray}
\mu(t _{i})&=&-\hbar {\partial_t}\theta(t _{i})-
J{\overline{v}^{2}_{s}(t_{i})}-V(t_{i}),\label{cp}
\end{eqnarray}

\noindent If we  make a change of variables $(t_i)\to
(R+(r/2),T+(t/2))$ in the one-point  functions $n_{o}(t_i)$, $\mu
(t_i)$, $v_s (t_i)$ and $V (t_i)$ and use the the slowly varying
dependence of the functions on the center of mass coordinates, to a
good approximation the functions can be treated as  continuous
functions  and  second order variations in $R$ and $T$  can  be
neglected. Thus, they  can be written as:
\begin{widetext}
\begin{eqnarray}
n_{o}(t_i) &=&
n_{o}(R+(r/2),T+(t/2))=n_{o}(R,T)+\frac{r}{2}\partial_R
n_{o}(R,T)+\frac{t}{2}\partial_T n_{o}(R,T), \label{apn}\\
\mu(t_i) &=& \mu(R+(r/2),T+(t/2))=\mu(R,T)+\frac{r}{2}\partial_R
\mu(R,T)+\frac{t}{2}\partial_T \mu(R,T),\label{apu}\\
v_s(t_i) &=& v_s(R+(r/2),T+(t/2))=v_s(R,T)+\frac{r}{2}\partial_R
v_s(R,T)+\frac{t}{2}\partial_T v_s(R,T),\label{apv}\\
V(t_i) &=&V(R+(r/2),T+(t/2))=V(R,T)+\frac{r}{2}\partial_R
V(R,T)+\frac{t}{2}\partial_T V(R,T).\label{apV}
\end{eqnarray}
\end{widetext}
\noindent Similar approximations can be made on the
two point functions by introducing  a change of variables
$(t_i,t_j)\to (r,\tau;R,T)$. The slowly varying dependence in $R$
and $T$ allow us to treat $\tilde{g}^{(\gtrless )}(r,\tau;R,T)$ as
a continuous functions  in the center of mass coordinates and
neglect second order variations in them. On the other hand, it is
important to include the discrete dependence  on the $r=i-j$
variables, inherent to the tight binding Hamiltonian, in order to
retain all the quantum effects introduced by the lattice which are
crucial to a proper description  of the system.

We now introduce a Fourier transform with respect to the relative
coordinate variables. Since hereafter we use the gauge-transformed
functions exclusively, the primes will be dropped to simplify the
notation:

\begin{eqnarray}
&&g^{(\gtrless )}(t_i,t_j^{\prime }) =g^{(\gtrless )}(r\tau;RT)
\label{fotra2}
\\&&\equiv-i \frac{1}{2\pi M}%
\sum_{q}\int d\omega e^{ \left( iqa_lr-i\omega \tau \right)} g^
{(\gtrless
)}(R,q;T,\omega ),\notag\\
&&H(t_i,t_j^{\prime }) =H(r\tau;RT)\label{fotrah}\\&&\equiv-i\frac{1}{2\pi M}%
\sum_{q}\int d\omega e^{ \left( iqa_lr-i\omega \tau \right)}\notag
H(R,q;T,\omega ),
\end{eqnarray}

\noindent using Eq. (\ref{apn}) in Eq. (\ref{fotrah}) we get:

\begin{equation}
H(R,q;T,\omega )=2 \pi M \left(I+\sigma_x \right){n_o}(R,T)\delta
(\omega )\delta _{q0}. \label{zero}
\end{equation}
\noindent In Eq.(\ref{zero}), the quantity $ {n_o}(R,T)$ is just
related to the condensate density of atoms at the space time point
$(Ra_l,T)$.  In Eq.(\ref{fotra2}), the upper diagonal component of
the two-point function ${g^ {<}_{11}(R,q;T,\omega )} $ corresponds
to the well known \emph{Wigner distribution} function
\cite{Wigner}. It can be interpreted as the density of
noncondensed particles with quasimomentum $q$ and energy
$\hbar\omega$ at the position  $Ra_l$ and time $T$. On the other
hand,  ${g^ {>}_{11}(qR;T,\omega )} $ is essentially the
\emph{density of states} available to a particle that is added to
the system at $(Ra_l,T)$ with quasimomentum $q$ and energy $\hbar
\omega$. As opposed to a normal system, the presence of the
condensate gives nonzero values to the off-diagonal terms of the
functions $ g_{12}^ {(\gtrless )}(R,q;T,\omega )$. We refer to
them as the \emph{anomalous contributions} to the respective two
point functions.

\subsection{Generalized  Boltzmann equations} \label{bolt22}

The generalized Boltzmann equations can be obtained \ as the
Fourier transform of the equations of motion \ for the case in
which the variations \ in $R$ and $T$ are very small: in
particular when the inverse propagator $D^{-1}_o$ and the self
energies vary very little as $Ra_l$ is changed by a characteristic
excitation wavelength or $T$ is changed by an inverse excitation
energy.

If we neglect the second order variation in $T$ and $R$, as
explained above,  the equations of motion Eq.(\ref{meankb}) to
(\ref{last}) can be approximated by:
\begin{widetext}
\begin{eqnarray}
\left( D_{o}^{-1}-\Re S+\frac{i}{2}\gamma\right) H
&=&-\frac{i}{2}\left[ D_{o}^{-1},H\right] +%
\frac{i}{2}\left[ \Re S,H\right] +\frac{1}{4}%
\left[ \gamma,H\right],  \label{ekbc1}\\
H\left( D_{o}^{-1}-\Re S-\frac{i}{2}\gamma\right)
&=&-\frac{i}{2}\left[H, D_{o}^{-1}\right] +%
\frac{i}{2}\left[H, \Re S\right] -\frac{1}{4}%
\left[ H,\gamma\right], \label{ekbc2}
\end{eqnarray}
\begin{eqnarray}
\left( D_{o}^{-1}-\Re \Sigma+\frac{i}{2}\Gamma\right) g^{(\gtrless
)}-\Sigma {}^{(\gtrless )}\left( \Re g+\frac{i}{2}A\right)
&=&-\frac{i}{2}\left[ D_{o}^{-1},g^{(\gtrless )}\right] +\frac{%
i}{2}\left[ \Re \Sigma ,g^{(\gtrless )}\right] + \frac{i}{2}\left[
\Sigma^{(\gtrless )},\Re g\right]\notag\\&&
+\frac{1}{4}\left[ \Gamma,g^{(\gtrless )}\right] -%
\frac{1}{4}\left[ \Sigma^{(\gtrless )},A\right], \label{e2}\\
g^{(\gtrless )}\left( D_{o}^{-1}-\Re\Sigma -\frac{i}{2}%
\Gamma\right) -\left( \Re g-\frac{i}{2}%
A\right) \Sigma {}^{(\gtrless )}
&=&-\frac{i}{2}\left[ g^{(\gtrless )},D_{o}^{-1}\right] +\frac{i%
}{2}\left[ g^{(\gtrless )},\Re \Sigma \right] + \frac{i}{2}\left[
\Re g,\Sigma^{(\gtrless )}\right]\notag\\&&
-\frac{1}{4}\left[ g^{(\gtrless )},\Gamma\right] +%
\frac{1}{4}\left[ A,\Sigma^{(\gtrless )}\right], \label{e3}
\end{eqnarray}
 \noindent with

\begin{equation}
D_{o}^{-1}(qR;T,\omega )\equiv \left(  \sigma_z\left( \hbar \omega
-\overline{v}_{s}(R,T)2J\sin (qa_l)\right) +\left( 2J\cos
(qa_l)+\mu (R,T)\right) I\right). \label{freep}
\end{equation}

\end{widetext}
\noindent In  Eqs. (\ref{ekbc1}-\ref{e3})all the quantities
depend on $(qR;T\omega)$.

\noindent In the equations we have also introduced the following
functions:

\begin{eqnarray}
\Re S (R,q;T,\omega )&=&S{^{HF}}(R,q;T,\omega )+\notag\\&&\Re
S^{B}(R,q;T,\omega ), \\
\Re \Sigma(R,q;T,\omega ) &=&\Sigma{^{HF}}(R,q;T,\omega
)+\notag\\&&\Re \Sigma^{B}(R,q;T,\omega ),
\end{eqnarray}
\begin{eqnarray}
&&\Re S^{B}(R,q;T,\omega ) =P\int \frac{d\omega ^{\prime }}{2\pi }\frac{%
\gamma(R,q;T,\omega )}{\omega -\omega ^{\prime }}, \\
&&\Re \Sigma ^{B}(R,q;T,\omega )=P\int \frac{d\omega ^{\prime }}{2\pi }%
\frac{ \Gamma(qR;T,\omega ^{\prime })}{\omega -\omega ^{\prime
}}, \\
&&\Re g(R,q;T,\omega ) =P\int \frac{d\omega ^{\prime }}{2\pi }%
\frac{A(R,q;T,\omega^{\prime })}{\omega -\omega ^{\prime }}.
\label{realg}
\end{eqnarray}

\noindent with $P$ denoting the Cauchy principal value and
$\gamma(R,q;T,\omega )$,  $\Gamma(R,q;T,\omega )$,
$S^{HF}(R,q;T,\omega )$,
$\Sigma^{HF}(R,q;T,\omega )$ and $%
A(R,q;T,\omega )$ understood as  Fourier transforms of the
functions $\gamma(t_i,t_j^{\prime })$,  $\Gamma(t_i,t_j^{\prime
})$, $S^{^{HF}}(t_i,t_j^{\prime })$, $\Sigma ^{HF}(t_i,t_j^{\prime
})$ and $A(t_i,t_j^{\prime })$ respectively.

To approximate the discretized equations by the continuous
differential equations  we have also used the slowly varying
dependence of the quantities on $R$ and $T$ . The brackets in
Eqs. (\ref{ekbc1}-\ref{e3}) denote the generalized Poisson
brackets defined as:

\begin{equation}
\lbrack A,B]=\frac{\partial A}{\partial \omega }\frac{\partial
B}{\partial
T}%
-\frac{\partial A}{\partial T}\frac{\partial B}{\partial \omega
}+\partial _{R}A\partial_{q}B-\partial_{q}A\partial _{R}B.
\end{equation}

\noindent Notice that even though the continuous limit has been
taken at the kinetic scale, the discreteness introduced by the
lattice, crucial for a correct description of the physics,  is
taken into account at the quantum scale,  as can be seen in Eq. (
\ref{freep}) where  the free propagator has a trigonometric
dependence on the quasimomentum $q$, characteristic of
lattice-type systems. If the disturbances in the system are small
enough that only long wavelength modes are excited, $qa\ll1$, the
excitations only see the lower quarter of the band. In this case
the free propagator reduces to

\begin{eqnarray}
&&D_{o}^{-1}((q\ll1/a_l)R;T,\omega )\approx  \sigma_z\left( \hbar
\omega -v_{s}(R,T)p\right)+\notag\\&&\left(
2J-\frac{p^2}{2m^{*}}+\mu(R,T) \right) I.
\end{eqnarray}

\noindent which is like the free propagator for a non-lattice system
and the role of the lattice is just to introduce an effective mass
$m^*$. Here $p=\hbar q$ and $m^*=\hbar^2/{(2a_l^2J)}$.

If we define the statistical functions(which carry a superscript
$^{(F)}$ in Paper I, Eq. (I.35)) as:
\begin{eqnarray}
F(R,q;T,\omega ) &=&\frac{g^{>}(R,q;T,\omega )+g^{<}(R,q;T,\omega )}{2}, \\
\Pi(R,q;T,\omega ) &=&\frac{\Sigma ^{>}(R,q;T,\omega )+\Sigma
^{<}(R,q;T,\omega )}{2},
\end{eqnarray}

\noindent  Eq. (\ref{e2})and   Eq. (\ref{e3}),  can  be rewritten
in terms of statistical and spectral functions as:

\begin{widetext}

\begin{eqnarray}
\left( D_{o}^{-1}-\Re \Sigma+\frac{i}{2}\Gamma\right) F-\Pi\left(
\Re g+\frac{i}{2}A\right)& =& \label{eqkb1}
-\frac{i}{2}\left\{\left[ D_{o}^{-1}- \Re \Sigma
+\frac{i}{2}\Gamma,F\right]
-\left[ \Pi,\Re g+ \frac{i}{2}A\right]\right\},\\
F\left( D_{o}^{-1}-\Re\Sigma -\frac{i}{2} \Gamma\right) -\left(
\Re g-\frac{i}{2} A\right)\Pi &=& \label{eqkb2}
-\frac{i}{2}\left\{\left[
F,D_{o}^{-1}-\Re \Sigma -\frac{i}{2}%
\Gamma\right] -\left[ \Re g-\frac{i}{2}A,\Pi \right]\right\},
\end{eqnarray}
\begin{eqnarray}
\left( D_{o}^{-1}-\Re \Sigma\right) A-\Gamma \Re g
&=&-\frac{i}{2}\left\{\left[ D_{o}^{-1}-\Re \Sigma,A\right]
-\left[ \Gamma,\Re g\right]\right\}\label{eqkb1h},\\
A\left(D_{o}^{-1}-\Re\Sigma\right)-\Re g\Gamma
&=&-\frac{i}{2}\left\{\left[ A,D_{o}^{-1}-\Re \Sigma\right]
-\left[ \Re g,\Gamma\right]\right\}. \label{eqkb2h}
\end{eqnarray}
\end{widetext}
 \noindent Eqs. (\ref{ekbc1}), (\ref{ekbc2}) and
(\ref{eqkb1})- (\ref{eqkb2h}) are our passage to the \ Boltzmann
equations. They describe the state of the gas at a given time.
Different from the HFB equations they include collisional
integrals for binary interactions.

\subsection{Ordinary  Boltzmann equations} \label{ordi}

To progress further we can  introduce  more simplifications based
on  physical considerations. The ordinary Boltzmann equation
emerges from the approximation in which the self energies that
appear on the left side of Eqs.(\ref{ekbc1}), (\ref{ekbc2}) and
(\ref{eqkb1})- (\ref{eqkb2h}) are handled differently from those
which appear on the right. These two appearance of the self-energy
play a different physical role in the description of the dynamics
\cite{Kadanoff}. The self energies on the right hand side describe
the dynamical effects of collisions, i.e., how the collisions
transfer particles from one energy-momenta configuration to
another.  On the other hand, the self energies on the left
describe the quantum kinetic effects due to interactions, i.e. how
interaction effects change the energy momentum  dispersion
relations from that of  free particles to a more complicated
spectrum. Because these two effects are physically distinct, we
can treat the left and the  right hand sides in a different way.

In the derivation of the ordinary Boltzmann equations, one
completely neglects  all the kinetic effects in the second order
self energies (the dependence on $T$ and $R$ in the  second order
self-energy terms on the right hand side) and retain dynamical
effects ($T$ and $R$ dependence on the left hand side). In this
way, we get the familiar Boltzmann equations which describe the
particles as free particles in between collisions with a modified
energy-momentum dispersion relation. It is a reasonable assumption
in dilute weakly interacting gases in which the duration of a
collision is very short compared to the essentially
interaction-free dynamics between isolated collisions.
 Neglecting kinetic effects in the second order self energies,
Eqs.(\ref{ekbc1}), (\ref{ekbc2}) and (\ref{eqkb1})- (\ref{eqkb2h})
can be approximated to

\begin{widetext}

\begin{eqnarray}
&&\left( D_{o}^{-1}-\Re S+\frac{i}{2}\gamma\right) H
=-\frac{i}{2}\left[ D_{o}^{-1}- S^{HFB},H\right],\\
&&H\left( D_{o}^{-1}-\Re S-\frac{i}{2}\gamma\right)
=-\frac{i}{2}\left[H, D_{o}^{-1}-S^{HFB}\right],\\
&&\left( D_{o}^{-1}-\Re \Sigma+\frac{i}{2}\Gamma\right)
F-\Pi\left( \Re g+\frac{i}{2}A\right) =-\frac{i}{2}\left[
D_{o}^{-1}- \Sigma^{HFB},F\right],\\
&&\left( D_{o}^{-1}-\Re \Sigma\right) A-\Gamma \Re
g=-\frac{i}{2}\left[
D_{o}^{-1}-\Sigma^{HFB},A\right],\\
&&F\left( D_{o}^{-1}-\Re\Sigma -\frac{i}{2}%
\Gamma\right) -\left( \Re g-\frac{i}{2}%
A\right)\Pi=-\frac{i}{2}\left[ F,D_{o}^{-1}-\Sigma^{HFB}\right], \\
&&A\left(D_{o}^{-1}-\Re\Sigma\right)-\Re g\Gamma
=-\frac{i}{2}\left[ A,D_{o}^{-1}-\Sigma^{HFB}\right].
\end{eqnarray}
\end{widetext}

If we take the trace of the sum and the  difference of each one of
the above equations with its hermitian conjugate, they can be
simplified to :

\begin{eqnarray}
Tr\left\{\left( D_{o}^{-1}-\Re S\right) H\right\}&=&0 \label{eqkin1},\\
Tr\left\{ \left(D_{o}^{-1}-\Re \Sigma\right) F-\Pi \Re g\right\}&=&0,\\
Tr\left\{ \left(D_{o}^{-1}-\Re \Sigma\right)A-\Gamma \Re
g\right\}&=&0,
\label{eqkin3}\\
Tr \left[ D_{o}^{-1}- S^{HFB},H\right]&=&-Tr (\gamma H),  \label{eqkin4}\\
Tr \left[ D_{o}^{-1}-\Sigma^{HFB},F \right]&=&-Tr(\Gamma  F-\Pi
A),\label{eqkin5}\\
Tr \left[ D_{o}^{-1}-\Sigma^{HFB},A\right]&=&0. \label{eqkin6}
\end{eqnarray}

Moreover, if we define  the operator $\daleth r M=M_{12}+M_{21}^*$
and apply it again to the sum and the difference of each one of
the equations (\ref{eqkb1}) to  (\ref{eqkb2h})  with its transpose
we also get:
\begin{widetext}
\begin{eqnarray}
Re\left(\daleth r\left\{\left( D_{o}^{-1}-\Re
S\right)H\right\}\right)&=& \frac{1}{2}Im\left(\daleth r \left[
D_{o}^{-1}- S^{HFB},H\right]+\daleth(\gamma H)\right),\label{eqkin1ap}\\
 Re\left(\daleth r\left\{ \left(D_{o}^{-1}-\Re \Sigma\right) F-\Pi \Re
g\right\}\right)&=&\frac{1}{2}Im\left(\daleth r \left[
D_{o}^{-1}-\Sigma^{HFB},F \right]+\daleth r(\Gamma  F-\Pi
A)\right),
\label{eqkin1a}\\
Re\left(\daleth r\left\{ \left(D_{o}^{-1}-\Re
\Sigma\right)A-\Gamma \Re g\right\}\right)&=&
\frac{1}{2}Im\left(\daleth r \left[
D_{o}^{-1}-\Sigma^{HFB},A\right]\right),
\label{eqkin3a}\\
Im\left(\daleth r\left\{\left( D_{o}^{-1}-\Re
S\right)H\right\}\right)&=& -\frac{1}{2}Re\left(\daleth r \left[
D_{o}^{-1}- S^{HFB},H\right]+\daleth(\gamma H)\right), \label{eqkin4a}\\
Im\left(\daleth r\left\{ \left(D_{o}^{-1}-\Re \Sigma\right) F-\Pi
\Re g\right\}\right)&=& -\frac{1}{2}Re\left(\daleth r \left[
D_{o}^{-1}-\Sigma^{HFB},F
\right]+\daleth r(\Gamma  F-\Pi A)\right),  \label{eqkin5a}\\
Im\left(\daleth r\left\{ \left(D_{o}^{-1}-\Re
\Sigma\right)A-\Gamma \Re g\right\}\right)&=&
-\frac{1}{2}Re\left(\daleth r \left[
D_{o}^{-1}-\Sigma^{HFB},A\right]\right),\label{eqkinf}
\end{eqnarray}

\end{widetext}
\noindent with $Re$ and $Im$ denoting the real and
imaginary parts. To close the set of equations, we need an
equation of motion for the superfluid velocity which can be found
from the definitions Eq.(\ref{cp}) and Eq.(\ref{suv}) to be:

\begin{equation}
\frac{\partial \overline{v}_{s}(R,T)}{\partial T}=-\frac{\partial
}{\partial R}\left( (\mu(R,T) + V(R,T) )+ J
{\overline{v}_{s}^{2}(R,T)}\right). \label{close}
\end{equation}

Eqs.(\ref{eqkin1}-\ref{eqkinf}) together with Eq.(\ref{close})
form a closed set of equations that describe the state of the gas
at a given time. Equations (\ref{eqkin1}-\ref{eqkin3}) and
(\ref{eqkin1ap}-\ref{eqkin3a}) are usually called the \emph{gap
equations}. They describe \ the quantum properties of a gas which
is evolving according to  Boltmaznn-type  equations
(\ref{eqkin4}-\ref{eqkin6}) and (\ref{eqkin4a}-\ref{eqkinf}).
Under the derived formalism Eqs.(\ref{eqkin1})-(\ref{close}) form
\ a coupled set of equations which \ replace the original
dynamics. The equations have to be solved \ self consistently for
any analysis.

\section{Equilibrium properties for a  homogeneous system}
\label{sec5}

 There are two situations in
which we expect an equilibrium solution to come from the Boltzmann
equations. Firstly when the system has never been disturbed and it
remains in its equilibrium state. Secondly when the system has had
sufficient time to relax after an applied perturbation. In this
section we will show how  the second order nonequilibrium
Boltzmann Equations lead, in these special cases  to the  linear
equilibrium solutions obtained from the HFB approximation
\cite{Bogoliubov} upgraded with second order corrections in $U$.

At equilibrium, in the absence of any external potential, the
functions $g^{\gtrless}$ and $H$ are completely independent of $R$
and $T$. In this case  the generalized Poisson-bracket terms are
zero and  Eqs (\ref{eqkin3}, \ref{eqkin6}) and (\ref{eqkin3a},
\ref{eqkin6}) imply that:

\begin{eqnarray}
A\left(D_{o}^{-1}-\Re\Sigma\right)-(\Re g)\Gamma&=&0.
\label{equaA}
\end{eqnarray}

\noindent Because $\Re g(q,\omega)$ is determined by $A(q,\omega)$
as indicated in Eq. (\ref{realg}),   Eq. (\ref{equaA}) is
satisfied when $A(q,\omega)$  is given by

\begin{eqnarray}
-i A(q,\omega)=\left[D_{o}^{-1}-\Re
\Sigma+\frac{i}{2}\Gamma\right]^{-1}- \left[D_{o}^{-1}-\Re
\Sigma-\frac{i}{2}\Gamma\right]^{-1} \label{decay}
\end{eqnarray}
\noindent and the function $\Re g(q,\omega)$ given by
\begin{eqnarray}
&&\Re g(q,\omega)=P\int \frac{d\omega^\prime}{2 \pi}
\frac{A(q,\omega^\prime)}{\omega-\omega^{\prime}}\\&&=
\frac{1}{2}\left\{\left[D_{o}^{-1}-\Re
\Sigma+\frac{i}{2}\Gamma\right]^{-1}+\left[D_{o}^{-1}-\Re
\Sigma-\frac{i}{2}\Gamma\right]^{-1}\right\}.\notag
\end{eqnarray}

\noindent From Eqs.(\ref{eqkin4}), (\ref{eqkin4a}), (\ref{eqkin5})
and (\ref{eqkin5a})  we also get, at equilibrium, the conditions
\begin{eqnarray}
\gamma&=&0,  \label{balc}\\
\Gamma F- \Pi A&=&0. \label{baf}
\end{eqnarray}

\noindent Eqs. (\ref{balc}) and (\ref{baf}) \ are just the
mathematical statement of detailed balance. They represent the
physical condition that at equilibrium the net rate of change \ of
the density of particles with momentum $q$ and energy $\omega $ \
is zero. Since it is always possible to write \cite{Kadanoff}

\begin{eqnarray}
F(q,\omega )&=&\left( n_q({\omega})+\frac{1}{2}\right)A(q,\omega),
\label{relat}
\end{eqnarray}
\noindent Eq.(\ref{baf}) can only be satisfied if

\begin{eqnarray}
\Pi(q,\omega )&=& \left(
n_q({\omega})+\frac{1}{2}\right)\Gamma(q,\omega),
\end{eqnarray}

\noindent is satisfied.  Detailed study of the structure of the
self-energy  indicates that $n_q(\omega)$ is  related to  the
Bose-Einstein thermal distribution, $n_q(\omega)=\frac{1}{e^{\beta
\omega }-1}$ with $\beta $ interpreted as  the local inverse
temperature in energy units \cite{Kadanoff,Kane}. In refs.
\cite{thermal} the authors prove that the only translational
invariant solution is the thermal.

 \noindent Since $H$ contains  delta functions in momentum and
energy at equilibrium, we  get from Eq. (\ref{eqkin1}):

\begin{eqnarray}
\mu&=& -2J+\Re S_{11}(0,0)+ \Re S_{12}(0,0). \label{chem}
\end{eqnarray}

\subsection{Quasiparticle formalism}
In the noninteracting case the diagonal terms of $A(q,\omega)$ are
just delta functions  with peaks at values of $\hbar\omega$ that
match the possible energy difference which results from adding a
single particle with quasimomentum $q$ to the system. In the many
body system the energy spectrum is sufficiently complex so  that
the diagonal elements of $A(q,\omega)$ are not delta functions but
instead  continuous functions of $\omega$. However, there are
always sharp peaks in $A$.  These sharp peaks represent the
coherent and long lived  excitations  which behave like weakly
interacting  particles.  These excitations are called
quasiparticles.  From Eq. (\ref{decay}) it is possible to see that
the quasiparticle decay rate is determined by $\Gamma$. The
quasiparticle approximation is obtained by considering $\Gamma $
very small for small values of $\omega$. This assumption  implies
that $D^{-1}\equiv D_{o}^{-1}-\Re \Sigma - \frac{i}{2}\Gamma$ is
essentially real with only an infinitesimal \ imaginary part. The
zeros of $D^{-1}$ about which $A$ is very sharply peaked are
identified with the quasiparticle energies $\hbar \omega_{q}$.

Using   the assumption of a very small $\Gamma$, and the identity
\begin{equation}
\lim_{\epsilon \to 0} \frac{1}{\omega -\omega ^{\prime }+i\epsilon
}=P\frac{1}{\omega -\omega ^{\prime }}-i\pi\delta(\omega -\omega
^{\prime }), \label{iden}
 \end{equation}

\noindent it is possible  to
 write  the  matrix components of $D^{-1}$  as:

\begin{eqnarray}
&&D^{-1}(q,\omega)= \\\notag&&\hbar \omega \left(
\begin{array}{cc}
 1 & 0 \\
0 & -1
\end{array}
\right) -\left(
\begin{array}{cc}
\mathcal{L}_{qq}(q,\omega) & \mathcal{M}_{q-q}(q,\omega) \\
\mathcal{M}_{q-q}^{\ast }(-q,-\omega) & \mathcal{L}_{qq}^{\ast
}(-q,-\omega)
\end{array}
\right),
\end{eqnarray}
\noindent with
\begin{eqnarray}
\mathcal{L}_{qq}(\omega )& =&-2J\cos qa_l-\mu +\Sigma
_{11}^{HFB}(q,\omega)\notag\\&&\label{quasi}+\int \frac{%
d\omega ^{\prime }}{2\pi }\frac{\Gamma_{11}(q,\omega ^{\prime })}{%
\omega -\omega ^{\prime }+i\epsilon }  , \\
\mathcal{M}_{q-q}(\omega ) &=&\Sigma  _{12}^{HFB}(q,\omega)+\int
\frac{d\omega ^{\prime }}{%
2\pi }\frac{\Gamma_{12}(q,\omega ^{\prime })}{\omega -\omega
^{\prime }+i\epsilon }. \label{quasian}
\end{eqnarray}

\noindent The quasiparticle amplitudes $u_q$ and $v_q$ are the
solutions to the eigenvalue problem
\begin{equation}
 \left(
\begin{array}{cc}
\mathcal{L}_{qq}(q,\omega_{q}) & \mathcal{M}_{q-q}(q,\omega_{q})
\\\mathcal{M}_{q-q}^{\ast }(-q,-\omega_{q}) & \mathcal{L}_{qq}^{\ast
}(-q,-\omega_{q})
\end{array}
\right)\left(
\begin{array}{c}
u_q \\
v_q
\end{array}
\right)=\hbar \omega_q \left( \begin{array}{c}
u_q \\
-v_q
\end{array}
\right)\label{quasift2} ,
\end{equation}
\noindent and satisfy the normalization condition
$|u_{q}|^{2}-|v_{q}|^{2}=1$. In the absence of vortices it is
always possible to find \ an ensemble \ in which the amplitudes
$(u_{q},v_{q})$ are purely real and  $u_{q}=u_{-q},\quad
v_{q}=v_{-q} $. In terms of  the quasiparticle amplitudes, the
matrix elements of the spectral function $A$, Eq.(\ref{decay}),
are given by:

\begin{eqnarray}
A_{11}(q,\omega ) &=&-2
Im\left[\frac{u_q^2}{\omega-\omega_q+i0^+}-\frac{v_q^2}{\omega-\omega_q+i0^-
}\right]\notag\\
&=&2  \pi \left[u_{q}^{2}\delta (\omega -\omega_{q})
-v_{q}^{2}\delta (\omega +\omega_{q})\right],
\\A_{12}(q,\omega ) &=&2 Im\left[\frac{u_q v_q
}{\omega-\omega_q+i0^+}-\frac{v_q
u_q}{\omega-\omega_q+i0^-}\right]\notag\\&=&-2\pi u_q v_q \left[
\delta (\omega -\omega_{q}) -
\delta (\omega +\omega_{q})\right],  \\
A_{22}(q,\omega ) &=&-A_{11}(-q,-\omega ),\\
A_{21}(q,\omega ) &=&A^{\ast}_{12}(q,\omega ).\end{eqnarray}

Finally, using the  definitions of $F$ and $A$, we can express the
matrix components: $\rho_q(\omega)$, $\widetilde{\rho}_q(\omega)$
and $m_q(\omega)$ defined as the Fourier transform of $\rho_{ij}$,
$\widetilde{\rho}_{ij}$ and $ m_{ij}$ respectively (see Eqs.
(\ref{defmat1}) and (\ref{defmat2})) in terms of quasiparticle
amplitudes:

\begin{eqnarray}
\rho_q(\omega) &=&2\pi \left[u_{q}^{2}n_q(\omega)\delta (\omega
-\omega_{q}) \right.
\\\notag &&\left.+v_{q}^{2}(1+n_q(\omega))\delta (\omega
+\omega_{q})\right],\\
\widetilde{\rho}_q(\omega) &=&2\pi
\left[u_{q}^{2}(1+n_q)(\omega)\delta (\omega -\omega_{q})
 \right.
\\&&  \notag \left.+v_{q}^{2}n_q(\omega)\delta (\omega +\omega_{q})\right],
\\m_q(\omega ) &=&2\pi u_q v_q
\left[ n_q(\omega)\delta (\omega +\omega_{q})  \right.
\\&& \notag \left.-
(1+n_q(\omega))\delta (\omega +\omega_{q})\right].
\end{eqnarray}

\subsection{HFB  approximation}

Under the HFB approximation the matrix $\ \Re \Sigma $ $\ $\ and
$\Re S$ are just given by $\Sigma ^{HFB}$ and $S^{HFB}$. In terms
of the quasiparticle amplitudes and setting  $\mathcal{N}=2$,
they can be written as:

\begin{widetext}
\begin{eqnarray}
\Sigma ^{HFB} &=&U\left(
\begin{array}{cc}
2\left( n_o +\widetilde{n}\right) &  n_o+%
\widetilde{m} \\
n_o +\widetilde{m} & 2\left( n_o+%
\widetilde{n}\right)
\end{array}
\right), \\
S^{HFB} &=&U \left(
\begin{array}{cc}
n_o+2\widetilde{n} & \widetilde{m} \\
\widetilde{m} &  n_o+2\widetilde{n}
\end{array}
\right).
\end{eqnarray}
with
\begin{eqnarray}
\widetilde{n} &=&=\frac{1}{M}\sum_{q}\left[
(1+n_q(\omega_q))v_{q}^{2}+u_{q}^{2}
n_{q}\right], \\
\widetilde{m} &=&\frac{1}{M}\sum_{q}u_{q}v_{q}\left(
2n_{q}(\omega_q)+1\right).
\end{eqnarray}

\noindent In the HFB approximation, Eq.(\ref{quasift2}) and Eq.
(\ref{chem}) then yield:

\begin{equation}
\left(
\begin{array}{cc}
-2J\cos( qa_l)-\mu +2U(n_o+\widetilde{n}) & U\left( n_{o}+%
\widetilde{m}\right) \\
U\left( n_o+\widetilde{m}\right) & -2J\cos (qa_l)-\mu +2U(n_o
+\widetilde{n})
\end{array}
\right) \left(
\begin{array}{c}
u_{q} \\
v_{q}
\end{array}
\right) =\hbar \omega_{q}\left(
\begin{array}{c}
u_{q} \\
-v_{q}
\end{array}
\right),  \label{hqure}
\end{equation}

\begin{equation}
\mu =-2J+Un_o+2U\widetilde{n}+U\widetilde{m}. \label{hfbvre}
\end{equation}
\end{widetext}
\noindent As a final step, to fix the total density to $n$, the
constraint
\begin{equation}
n=n_o+\widetilde{n},\label{hfbcont}
\end{equation}
\noindent has to be satisfied.

For a given density and temperature Eqs. (\ref{hqure})-
(\ref{hfbcont}) form a closed set of equations. At zero
temperature, they reduce to the HFB equations derived in
\cite{Bogoliubov} using the quadratic  approximation.

\ The Hugenholtz-Pines theorem states\cite{Hugenholtz} that a
homogeneous system at equilibrium has to fulfill

\begin{equation}
\mathcal{L}_{qq}(0,0)-\mathcal{M}_{q-q}(0,0)=0
\end{equation}
The above equation  implies that the energy spectrum of a Bose gas
is gapless,i.e. there is an excitation with an energy that tends
to zero in the limit of zero momentum. Mathematically the theorem
implies that the two-point propagator $g(q,\omega )$ has a pole at
$q=\omega =0$. Physically it reflects the fact that small
rotations of the phase of the condensate wave function cost little
energy (Goltstone mode of the broken symmetry). The
Hugenholtz-Pines theorem is a consequence of the invariance of the
mean field and the two point propagators under a phase
transformation.

The HFB approximation  violates the Hugenholtz-Pines theorem:

\begin{equation}
\mathcal{L}_{qq}(0,0)-\mathcal{M}_{q-q}(0,0)=-2U\widetilde{
m}\neq0
\end{equation}

One way to solve the gap problem is to set  the anomalous term
$\widetilde{ m}$ to zero in HFB  equations. This procedure is
known as HFB-Popov approximation. The HFB-Popov equations were
first introduced by Popov \cite{Popov}, and at equilibrium they
are consider a better approximation than the HFB equations because
they yield a gapless spectrum. Nevertheless the HFB-Popov
equations are not conserving and therefore they are not
appropriate to describe dynamical evolution.

\subsection{Second-order and Beliaev  approximations}

\bigskip  When second order terms
are taken into account the matrices $\mathcal{L}_{qq}$ and
$\mathcal{M}_{q-q}$  become energy dependent. For simplicity we
restrict the calculations to the zero temperature case when
$n_q=0$. In terms of the quasiparticle amplitudes  the
contributions to the self-energy  at second order are given by
\begin{widetext}
\begin{eqnarray}
&&\mathcal{M}_{q-q}(q,\omega ) =Un_o+U\widetilde{m}+  \\ \notag
&&\frac{2 U^{2}}{\hbar M}n _{o}\sum_{k}\left(
\frac{2\mathcal{A}_{k}\mathcal{B}_{q-k}+2\mathcal{C}_{k}
\mathcal{A}_{p}+ 2\mathcal{C}_{k}\mathcal{B}_{q-k}+
3\mathcal{C}_{k}\mathcal{C}_{q-k} }{%
\omega -\omega_{k}-\omega_{q-k}+i\epsilon }\right. \left.
-\frac{2\mathcal{B}_{k}\mathcal{A}_{q-k}+2\mathcal{C}_{k}
\mathcal{A}_{p}+ 2\mathcal{C}_{k}\mathcal{B}_{q-k}+
3\mathcal{C}_{k}\mathcal{C}_{q-k}}{\omega
+\omega_{k}+\omega_{q-k}-i\epsilon
}%
\right) \\\notag &&+ \frac{2U^{2}}{\hbar M^2}\sum_{k,p} \left(
\frac{2\mathcal{A}_{k}\mathcal{B}_{p}\mathcal{C}_{q-k-p}+\mathcal{C}_{k}\mathcal{C}_{p}\mathcal{C}_{q-k-p}}
{\omega -\omega_{k}-\omega_{p}-\omega_{q-k-p}+i\epsilon
}-\frac{2\mathcal{B}_{k}\mathcal{A}_{p}\mathcal{C}_{q-k-p}+\mathcal{C}_{k}\mathcal{C}_{p}\mathcal{C}_{q-k-p}}{\omega
+\omega_{k}+\omega_{p}+\omega_{q-k-p}-i\epsilon }\right),\\
&&\mathcal{L}_{qq}(q,\omega ) =-2J\cos qa_l-\mu +
2Un_o+2U\widetilde{n}+ \\\notag &&
  \frac{2U^{2}n_o}{\hbar M}%
\sum_{k}\left( \frac{
\mathcal{A}_{k}\mathcal{A}_{q-k}+2\mathcal{A}_{k}\mathcal{B}_{q-k}+4\mathcal{C}_{k}\mathcal{A}_{q-k}+
2\mathcal{C}_{k}\mathcal{C}_{q-k}}{\omega
-\omega_{k}-\omega_{q-k}+i\epsilon } \right. - \left.
\frac{%
\mathcal{B}_{k}\mathcal{B}_{q-k}+2(\mathcal{B}_{k}\mathcal{A}_{q-k})+4\mathcal{C}_{k}\mathcal{B}_{q-k}+
2\mathcal{C}_{k}\mathcal{C}_{q-k}}{\omega
+\omega_{k}+\omega_{q-k}-i\epsilon }\right)\\\notag
&&+\frac{2U^{2}}{\hbar M^2}\sum_{k,p}\left( \frac{%
\mathcal{A}_{k}\mathcal{A}_{p}\mathcal{B}_{q-k-p}+2\mathcal{A}_{k}\mathcal{C
}_{p}\mathcal{C}_{q-k-p}}{\omega
-\omega_{k}-\omega_{p}-\omega_{q-k-p}+i\epsilon }\right) -\left( \frac{%
\mathcal{B}_{k}\mathcal{B}_{p}\mathcal{A}_{q-k-p}+2\mathcal{B}_{k}\mathcal{C
}_{p}\mathcal{C}_{q-k-p}}{\omega
+\omega_{k}+\omega_{p}+\omega_{q-k-p}-i\epsilon }\right),\\
&&\mu  =-2J+Un_o+2U\widetilde{n}+U\widetilde{m} -\frac{2
U^{2}}{\hbar M^2}\sum_{k,p}\left(
\frac{2\mathcal{A}_{k}\mathcal{B}_{p}\mathcal{C}_{k+p}+2\mathcal{B}_{k}\mathcal{A}_{p}
\mathcal{C}_{k+p}+2\mathcal{C}_{k}\mathcal{C}_{p}\mathcal{C}_{k+p}
}{\omega_{k}+\omega_{p}+\omega_{k+p}}\right) \notag \\ \notag
&&-\frac{2U^{2}}{\hbar M^2}\sum_{k,p}\left(
\frac{2\mathcal{A}_{k}\mathcal{C}_{p}\mathcal{C}_{k+p}+
\mathcal{A}_{k}\mathcal{A}_{p}\mathcal{B}_{k+p}+2\mathcal{B}_{k}\mathcal{C}_
{p}\mathcal{C}_{k+p}+
\mathcal{B}_{k}\mathcal{B}_{p}\mathcal{A}_{k+p}}{%
\omega_{k}+\omega_{p}+\omega_{k+p}}\right),
\end{eqnarray}
\end{widetext}

\noindent where the quantities $\mathcal{A}$, $\mathcal{B}$ and
$\mathcal{C}$ are defined as

\begin{eqnarray}
\mathcal{A}_{k} &=&u_{k}^{2}, \quad \mathcal{B}_{k}=v_{k}^{2},
\quad \mathcal{C}_{k} =-u_{k}v_{k}.
\end{eqnarray}

 The inclusion of second order terms modifies  the
structure of the HFB equations. The matrix that we need to
diagonalize to find the quasiparticle energies  depends now  on
the quasiparticle mode in consideration. This means that a
separate nonlinear problem must be solved for every quasiparticle
state, whereas the solution of the HFB equations yields the whole
quasiparticle spectrum. The matrix which is to be diagonalized
also becomes intrinsically nonlocal and to solve for a
quasiparticle state with quasimomentum $q$ we have to sum over
alldifferent quasimomenta. Finally, the diagonal elements  are no
longer equal as was always the case in all the quadratic
approximations.

If we omit the second order terms containing no condensate
amplitudes, the equations that we get are the \emph{tight-binding}
version of the ones originally derived by Beliaev \cite{Beliaev}:
\begin{eqnarray}
\mathcal{M}_{q-q}(q,\omega ) &=&U n_o+
\lambda \Delta\mathcal{M}_{q-q}(q,\omega )\\
\mathcal{L}_{qq}(q,\omega ) &=& \epsilon[q] +U n_o+ \lambda \Delta
\mathcal{L}_{qq}(q,\omega )\\
\mu  &=&-2J+U n_o + \lambda \Delta \mu.
\end{eqnarray}

\noindent with
\begin{eqnarray}
\epsilon[q]&=&4J\sin^2(qa_l/2)\\
 \Delta \mu &=&2U\widetilde{n}+U\widetilde{m}.
\end{eqnarray}
\noindent and
\begin{widetext}
\begin{eqnarray}
&&\Delta\mathcal{M}_{q-q}(q,\omega ) = U\widetilde{m}+ \frac{2
U^{2}}{\hbar M}n _{o}\sum_{k}\left(
\frac{2\mathcal{A}_{k}\mathcal{B}_{q-k}+2\mathcal{C}_{k}
\mathcal{A}_{p}+ 2\mathcal{C}_{k}\mathcal{B}_{q-k}+
3\mathcal{C}_{k}\mathcal{C}_{q-k} }{%
\omega -\omega_{k}-\omega_{q-k}+i\epsilon }\notag \right.\\&&-
\left. \frac{2\mathcal{B}_{k}\mathcal{A}_{q-k}+2\mathcal{C}_{k}
\mathcal{A}_{p}+ 2\mathcal{C}_{k}\mathcal{B}_{q-k}+
3\mathcal{C}_{k}\mathcal{C}_{q-k}}{\omega
+\omega_{k}+\omega_{q-k}-i\epsilon
}\right), \label{beliaev1}\\
&&\Delta\mathcal{L}_{qq}(q,\omega ) = - U\widetilde{m}+
\frac{2U^{2}n_o}{\hbar M}%
\sum_{k}\left( \frac{%
\mathcal{A}_{k}\mathcal{A}_{q-k}+2\mathcal{A}_{k}\mathcal{B}_{q-k}+4\mathcal
{C}_{k}\mathcal{A}_{q-k}+
2\mathcal{C}_{k}\mathcal{C}_{q-k}}{\omega
-\omega_{k}-\omega_{q-k}+i\epsilon } \right. \notag\\&&- \left.
\frac{\mathcal{B}_{k}\mathcal{B}_{q-k}+2(\mathcal{B}_{k}\mathcal{A}_{q-k})+4
\mathcal{C}_{k}\mathcal{B}_{q-k}+
2\mathcal{C}_{k}\mathcal{C}_{q-k}}{\omega
+\omega_{k}+\omega_{q-k}-i\epsilon
}\right),\label{beliaev2}\end{eqnarray}

\end{widetext}
 \noindent In the above equations we introduce the
parameter $\lambda$  only to use it  as a perturbation parameter
and set to one at the end of the calculations.

If second order terms are included in the theory, they change  the
quasiparticle  spectrum not only   by shifting the quasiparticle
energies but also by making them complex. The imaginary part that
the quasiparticle energies acquire comes from the poles of the
second order terms and it is associated with a damping rate. The
physical meaning  is that when the energy denominator in the
second order terms vanishes  a process where a quasiparticle
decays into two of lower energy is energetically allowed. This
kind of damping mechanism is known as \emph{Beliaev damping} and
was calculated by Beliaev in the case of a uniform Bose superfluid
\cite{Beliaev}. In the remainder of this section we calculate the
zero temperature Beliaev damping coefficient for atoms in optical
lattices using the tight-binding second order Beliaev
approximation, Eqs.~(\ref{beliaev1})-(\ref{beliaev2}). We follow
the same ideas used by  Beliaev  to study the uniform system.

\subsubsection{Perturbative treatment}

As the starting point we  assume that the net effect of second
order plus HFB terms is to introduce small corrections to the
Bogoliubov-de Gennes  (BdG) self energies \footnote{%
In the translationally invariant limit the Bogoliubov-de Gennes
matrix elements $\mathcal{L}_{qq}$ and $\mathcal{M}_{q-q}$  agree
with the matrix elements calculated using the HFB-Popov
\cite{Morgan,Bogoliubov} }. In this case instead of solving the
equations in a self consistent way we can replace the BdG
quasiparticle energies and amplitudes in the HFB and second order
self-energy corrections to calculate the shift they introduce in
the spectrum.

The quasiparticle energies and amplitudes in the BdG approximation
are given by \cite{Bogoliubov}:

\begin{equation}
\hbar \omega ^{(0)}_{q}=\sqrt{{\varepsilon_{q}}^2+ 2U
n_{o}^{(0)}\varepsilon_{q}}, \label{wbh}
\end{equation}

\begin{equation}
\mathcal{A}^{(0)}_{q}={u^{(0)}_{q}}^{2} =\frac{\varepsilon
_{q}+n_{o}^{(0)}U+\hbar\omega_{q}^{(0)}}{ 2 \hbar\omega
_{q}^{(0)}}\label{us},
\end{equation}
\begin{equation}
\mathcal{B}^{(0)}_{q}={v^{(0)}_{q}}^{2}=\frac{\varepsilon
_{q}+n_{o}^{(0)}U-\hbar\omega_{q}^{(0)}}{ 2\hbar \omega
_{q}^{(0)}}, \label{vs}
\end{equation}

\begin{equation}
\mathcal{C}^{(0)}_{q}=-u^{(0)}_{q}v^{(0)}_{q}=-\frac{n_{o}^{(0)}U
}{2\hbar\omega _{ q}^{(0)}}, \label{usvs}
\end{equation}

\begin{equation}
 \widetilde{m}^{(0)}= \frac{1}{M} \sum_{q\neq 0}
{u^{(0)}_{q}}{v^{(0)}_{q}},
\end{equation}

\noindent and

\begin{equation}
n= n_{o}^{(0)}+\frac{1}{M} \sum_{q\neq 0}
{v^{(0)}_{q}}^{2},\label{nucons}
\end{equation}

 \noindent with $n$ the total density, $n=N/M$.

As shown in the last section, the HFB approximation has the
problem that it has a gap in the excitation spectrum and therefore
violates Pines-Hugenholtz theorem. However, as shown by Beliaev
\cite{Beliaev}, when second order Beliaev  contributions are
included the theory becomes gapless.  This can be seen from
Eqs.(\ref{beliaev1}) and (\ref{beliaev2}):
\begin{widetext}
\begin{eqnarray}
\Delta\mathcal{L}_{q-q}(0,0)-\Delta\mathcal{M}_{q-q}(0,0)&=&-2U\widetilde{m}
^{(0)}+
 \frac{2U^{2}n_o}{\hbar M}%
\sum_{k}\left( \frac{%
\mathcal{A}^{(0)}_{k}\mathcal{A}^{(0)}_{-k}+\mathcal{B}^{(0)}_{k}\mathcal{B}
^{(0)}_{-k}-2\mathcal{C}^{(0)}_{k}\mathcal{C}^{(0)}_{-k} }{
-2\omega_{k}^{(0)}} \right)\notag\\&=&-2U\widetilde{m}^{(0)}-
\frac{2U^{2}n_o}{\hbar M}%
\sum_{k}\frac{\left({u_k^{(0)}}^2-{v_k^{(0)}}^2\right)^2 }{
-2\omega_{k}^{(0)}} \notag\\&=&2U\frac{1}{M}\sum_{k}\frac{U n_o}{
-2\hbar
\omega_{k}^{(0)}}-\frac{2U^{2}n_o}{\hbar M}%
\sum_{k}\frac{1}{ -2\omega_{k}^{(0)}}=0.
\end{eqnarray}
\end{widetext}
\subsubsection{Beliaev damping}

If we include HFB and second order corrections, the quasiparticle
energy shifts  are given to first order in $\lambda$ by
\begin{widetext}
\begin{eqnarray}
\hbar \omega ^{(1)}_{q}&\equiv&\delta E_q+ i\gamma_q \label{corr}
\\&=& \mathcal{A}^{(0)}_{q}\Delta\mathcal{L}_{q-q}(q,\omega
^{(0)}_{q})+\mathcal{B}^{(0)}_{q}\Delta\mathcal{L}_{q-q}^*(-q,-\omega
^{(0)}_{q}) +
\mathcal{C}^{(0)}_{q}(\Delta\mathcal{M}_{q-q}(q,\omega
^{(0)}_{q})+\Delta\mathcal{M}^*_{q-q}(-q,-\omega
^{(0)}_{q})).\notag
\end{eqnarray}
\end {widetext}
 \noindent  After some algebra, Eq.~(\ref{corr}) can
be written in the more convenient form:
\begin{widetext}
\begin{equation}
\delta E_q+ i\gamma_q=
U\widetilde{m}^{(0)}(u^{(0)}_{q}-v^{(0)}_{q})^2+
 \frac{4
U^{2}}{\hbar M}n _{o}^{(0)}\sum_{k}\left(
\frac{B_{k,q-k}^2 }{%
\omega^{(0)}_q -\omega^{(0)}_{k}-\omega^{(0)}_{q-k}+i\epsilon }-
\frac{\widetilde{B}_{k,q-k}^2 }{%
\omega^{(0)}_q +\omega^{(0)}_{k}+\omega^{(0)}_{q-k}-i\epsilon
}\right), \label{shift2}
\end{equation}

\noindent where the matrices $B_{k,q-k}$ and
$\widetilde{B}_{k,q-k}$ are defined as

\begin{eqnarray}
B_{k,q-k}&=&u^{(0)}_{q}(u^{(0)}_{k}u^{(0)}_{q-k}-u^{(0)}_{k}v^{(0)}_{q-k}-v^
{(0)}_{q}u^{(0)}_{q})-
v^{(0)}_{q}(v^{(0)}_{k}v^{(0)}_{q-k}-u^{(0)}_{k}v^{(0)}_{q-k}-v^{(0)}_{q}u^{
(0)}_{q}), \label{mat1}\\
\widetilde{B}_{k,q-k}&=&u^{(0)}_{q}(v^{(0)}_{k}v^{(0)}_{q-k}-u^{(0)}_{k}v^{(
0)}_{q-k}-v^{(0)}_{q}u^{(0)}_{q})-
v^{(0)}_{q}(u^{(0)}_{k}u^{(0)}_{q-k}-u^{(0)}_{k}v^{(0)}_{q-k}-v^{(0)}_{q}u^{
(0)}_{q}). \label{mat2}
\end{eqnarray}
\end{widetext}
\noindent If we replace $\varepsilon_{q}$ by $\hbar
q^2/2m$, the matrix elements given by  Eqs. (\ref{mat1}) and
(\ref{mat2}) reduce to the Beliaev uniform gas matrix elements
(see for example Ref. \cite{Morgan} and \cite{Giorgini}).

The damping coefficient  can be obtained using the identity Eq.
~(\ref{iden}) in Eq.~(\ref{shift2}). This yields

\begin{equation}
\gamma_q=  \frac{2 \pi U^{2}}{ M \hbar}n
_{o}^{(0)}\sum_{k}B_{k,q-k}^2\delta\left( \omega^{(0)}_q -
 \omega^{(0)}_{k}- \omega^{(0)}_{q-k}\right).
\label{bdam}
\end{equation}

\noindent For a translational-invariant system  at equilibrium,
all quantities are  $T$ and $R$ independent and depend only on the
relative coordinates $r$ and $\tau$. Therefore, at equilibrium the
scale separation is always valid and  we can relax the condition
$Un/J\ll1$. In Ref. \cite{Bogoliubov} we showed by comparison with
solutions obtained by  the exact diagonalization of the
Bose-Hubbard Hamiltonian,  that for commensurate systems in the
parameter regime where $(U/J)<0.5(U/J)_c$, the BdG equations give
a good description of the properties of the system. $(U/J)_c \sim
dn$ is the superfluid to Mott insulator critical ratio,  $d$ the
dimensionality  and $n$ the density of the system. For systems
with non-commensurate fillings, where the superfluid to Mott
insulator quantum phase transition does not take place, the
agreement between the BdG  and the exact solutions was shown to be
significantly better for a larger parameter regime. Because Eq.
(\ref{bdam}) was found treating the second order corrections  as a
perturbation, its validity  is restricted to the parameter regime
$(U/J)<0.5(U/J)_c$, where the BdG solutions are still a good
description of the system.

As opposed to the uniform system without the lattice, where for
high momentum  the single  particle energy ( which grows as
$q^2$)is always dominant, in the presence of the lattice, the
single particle excitation energies are always bounded by $4J$.
Therefore, in the regime  $Un/J>1$ \footnote{ Notice that for
large filling factors $n$, the parameter  $Un/J$ can be bigger
than one but the system can be still far away from the Mott
insulator critical point} the interaction term dominates for all
quasimomenta and the quasiparticle amplitudes and energies can be
expanded as:

\begin{eqnarray}
u^{(0)}_{k}&\simeq& \frac{1}{2\alpha_k} + \frac{\alpha_k}{2} +
\frac{{\alpha_k}^3}{8} -\frac{{\alpha_k}^5}{8}, \label{aus}\\
v^{(0)}_{k}&\simeq& \frac{1}{2\alpha_k} - \frac{\alpha_k}{2} +
\frac{{\alpha_k}^3}{8} +\frac{{\alpha_k}^5}{8}, \label{avs}\\
\hbar \omega^{(0)}_k&\simeq& 2n_o^{(0)}
U\left(\alpha_k^2+\frac{1}{2}\alpha_k^6\right), \label{aws}
\end{eqnarray}

\noindent where

\begin{eqnarray}
\alpha_k&=&\eta
\left(\frac{\varepsilon_k}{J}\right)^{1/4},\\
\eta&\equiv&\left(\frac{J}{2 n_o^{(0)} U}\right)^{1/4}.
\end{eqnarray}

\noindent In the very weakly interacting regime $Un_o/J\lesssim1$,
the approximations used to derive Eqs.~(\ref{aus}) to (\ref{aws})
are still valid if the quasimomentum of the excitation involved in
the decay process is small, $qa_l\ll\sqrt{n_o U/J }$.

If one substitutes  Eqs.~(\ref{aus}) to (\ref{aws}) for the
quasiparticle amplitudes  in Eq.(\ref{bdam}) and makes use of the
energy conservation condition, which is approximately given by

\begin{equation}
\alpha_q^2- \alpha_k^2 -\alpha_{q-k}^2 =\frac{1}{2}(\alpha_k^6
+\alpha_{q-k}^6-\alpha_q^6),
\end{equation}

\noindent one gets the following expression for the damping
coefficient:

\begin{equation}
\gamma_q= \frac{9 \pi}{8M}  \frac{ J}{ n
_{o}^{(0)}}\sum_{k}\sqrt{\frac { \varepsilon_q \varepsilon_k
\varepsilon_{q-k}}{J^3}}\delta\left( \overline{e}_q -
\overline{e}_{k}-\overline{e}_{q-k}\right), \label{beli}
\end{equation}

\noindent with $ \overline{e}_q$ the dimensionless quasiparticle
energies given by  $ \overline{e}_q= \frac{\hbar
\omega^{(0)}_{k}}{ 2 n _{o}^{(0)} U \eta^2}$. When  the number of
lattice sites is large, to a good approximation  the discrete sum
can be replaced by an integral $1/M \sum_{k}\to a_l/{2\pi}
\int_0^{2\pi/a_l} dk$.

 For the one dimensional system, we find that the only value of
 $k$ at which the energy constraint is satisfied is when  $k=q$.
This value of $k$ leads to a zero damping coefficient and
therefore in the one dimensional  system the   quasiparticles
become totally stable against their decay into two of lower
energy. In this case higher order decay processes have to be
considered. However, the absence of Beliaev damping  in one
dimensional lattices is not a particular characteristic of the
lattice dispersion relation. If the damping coefficient is
calculated using the  one dimensional uniform Bose gas dispersion
relation, it  is also found to be zero.

The extension of the expression for the Beliaev damping
coefficient to higher dimensional lattice systems can be done
straightforwardly. One just has to replace the single particle
dispersion relation $\varepsilon_k$ in Eq.\ref{beli} by the one in
the specific dimension. If we assume a  separable square lattice
in $d$ dimensions, with the same tunneling matrix energy $J$, and
lattice constant $a_l$,  in all different directions we get:

\begin{eqnarray}
\gamma_q ^{(d)}&=& \frac{9 \pi}{8M^d}  \frac{ J}{ n
_{o}^{(0)}}\sum_{\mathbf{k}}\sqrt{\frac { \varepsilon_\mathbf{q}
\varepsilon_\mathbf{k}
\varepsilon_{\mathbf{q}-\mathbf{k}}}{J^3}}\delta\left(
\overline{e}_\mathbf{q}-
\overline{e}_\mathbf{k}-\overline{e}_{\mathbf{q}-\mathbf{k}}\right)
\\\notag &\approx& \frac{9 a_l^d}{16(2 \pi)^{d-1}}  \frac{ J}{ n
_{o}^{(0)}}\int d {\mathbf{k}}\sqrt{\frac { \varepsilon_\mathbf{q}
\varepsilon_\mathbf{k}
\varepsilon_{\mathbf{q}-\mathbf{k}}}{J^3}}\delta\left(
\overline{e}_\mathbf{q}-
\overline{e}_\mathbf{k}-\overline{e}_{\mathbf{q}-\mathbf{k}}\right),
\end{eqnarray}

\noindent with the definitions $\varepsilon
_{\mathbf{k}}=4J\sum_{i=1}^{d}\sin
^{2}\left(\frac{k_{i}a_l}{2}\right)$, $\hbar
\omega^{(0)}_\mathbf{k}\simeq 2n_o^{(0)}
U\left(\alpha_\mathbf{k}^2+\frac{1}{2}\alpha_\mathbf{k}^6\right)$,
$\alpha_\mathbf{k}=\eta
\left(\frac{\varepsilon_\mathbf{k}}{J}\right)^{1/4}$ and
 $\overline{e}_{\mathbf{k}}= \frac{\hbar \omega^{(0)}_{\mathbf{k}}}{ 2 n
_{o}^{(0)} U \eta^2}$.

An analytic expression for the damping coefficient  can be easily
obtained when the excitations involved in the decay process have
long wave number: $qa_l\ll1$. In this parameter regime  for the
particular case of a three dimensional lattice  the integral
yields:

\begin{widetext}
\begin{eqnarray}
\gamma_{qa_l\ll1} ^{(d=3)} &\approx& \frac{9 }{32 \pi}  \frac{
Ja_l^3}{ n _{o}^{(0)}}\int dk d\theta \sin(\theta)k^2 \sqrt{\frac
{ \varepsilon_\mathbf{q} \varepsilon_\mathbf{k}
\varepsilon_{\mathbf{q}-\mathbf{k}}}{J^3}}\delta\left(
qa_l-ka_l-a_l\sqrt{p^2+q^2-2pq\cos\theta}\right)\notag
\\&\approx& \frac{9 }{32 \pi} \frac{ Ja_l^5}{ n
_{o}^{(0)}}\int_0^{qa_l} dk k^2(q-p)^2= \frac{3 }{640 \pi}  \frac{
\hbar^2 a_l^3 q^5}{ m^* n _{o}^{(0)}},
\end{eqnarray}
\end{widetext}
 \noindent with  $m^*=\hbar^2/(2Ja_l^2)$ the
effective mass. In the long wavelength limit, or phonon regime,
the damping coefficient in the lattice reduces to the well known
result first obtained by Beliaev in the phonon regime, with the
mass replaced by an effective mass.

Outside the phonon regime, the analytic evaluation  of the
integral is more complicated because of the energy conservation
constraint. In the uniform gas case, which has a simpler
quasiparticle spectrum, it has been shown that there is a finite
threshold momentum $q^*$ such that the decay  of an excitation is
impossible if $q>q^*$ \cite{Maris}. We expect that the
trigonometric dependence on the quasimomentum of the quasiparticle
dispersion relation in lattice-type systems  makes the energy
conservation constraint even harder to fulfill. In Ref.
\cite{Tsuchiya} the authors calculated the finite temperature
Landau damping coefficient  in a one dimensional optical lattice
and showed the disappearance of Landau damping when $Un_o/J\geq6$.

\section{Conclusions}
\label{sec6} In this work we continued our previous studies of
the dynamics of bosonic
 atoms confined in optical potentials. Here, starting from  the 2PI-CTP
equations of motion, derived in Paper I from  the Bose-Hubbard
Hamiltonian, we show how  the complicated nonlocal, non-Markovian
integro-differential equations can be simplified and reduced to
the standard kinetic theory equations. Specifically, by using a
two-time separation condition, valid in dilute weakly interacting
systems not very far away from equilibrium, we recast the full
quantum dynamics into two coupled sets of  equations: the first
set of Boltzmann equations governing the distribution functions
and  a second set of  gap equations describing the modified
dispersion relation. We conclude here with three  remarks on some
general features of this problem and our approach.

First, a remark on quantum kinetic theory in discrete versus
continuous systems: Even though we work with a lattice gas system
described by the Bose Hubbard Hamiltonian, the assumption that the
propagators are slowly varying in the center of mass coordinates
permits one to map the discrete tight binding equations into a set
of continuous differential equations in the center of mass
coordinates. For this reason the dynamical equations of motion we
derived for discrete systems look very similar to previous kinetic
equations derived for continuous systems. On the other hand,  to
include all the relevant dynamical effects introduced by the
lattice, we kept  the discrete character of the tight binding
Hamiltonian in the equations for the relative coordinates, as
manifested in the gap equations which exhibit a dispersion
relation different from the homogeneous Bose gas system.

Second, the last section of this work was dedicated to a study of
quantum equilibrium solutions. By using the quasiparticle
approximation, we recovered from the kinetic equations the linear
HFB corrections to the self-energy plus second order corrections.
We showed  how by neglecting the condensate-independent second
order terms in the self-energy, one obtains a tight-binding
version of the well known Beliaev equations. We used these
equations to derive expressions for the zero temperature Beliaev
damping coefficient in lattice systems in certain parameter
regimes. In particular, we showed that for long wavelength
excitations, the damping coefficient in a three dimensional
lattice reduces to the one calculated for a uniform Bose gas in
the phonon regime, but with the mass replaced by the effective
mass induced by the lattice.

A final remark on the purpose of this work. It is not meant to be
a mere academic exercise in our demonstration of how Boltzmann
like equations are obtained from the effective action and
equilibrium solutions can be obtained from the full quantal
solutions. In making explicit the simplifying assumptions en route
starting from first principles, it allow us to realize the
limitations and the applicability of a kinetic theory formulation
for describing the quantum dynamics of many-body lattice systems.
It serves to identify the range of validity  and the parameter
regimes where the underlying assumptions leading to these
simplified kinetic equations  can become unreliable. We view this
effort as having both theoretical and practical significance in
seeking a proper description of such systems and better
understanding of its behavior -- theoretical in scrutinizing the
practicing kinetic theories in existence, and practical in
providing the correct paramters for comparison with experiments.

\paragraph{Acknowledgement}
AMR and CW Clark are supported in part by an Advanced Research and
Development Activity (ARDA) contract and by the U.S. National
Science Foundation under grant PHY-0100767. BLH is supported in
part by  NSF grant PHY-0426696, a NIST grant and an ARDA contract
MDA90401/C0903. EC is supported by the University of Buenos Aires,
CONICET, fundacion Antorchas and ANPCyT under project PICT 99
03-05229.

\end{document}